\documentclass[reprint,amsmath,amssymb,aps,rmp]{revtex4-2}

\usepackage{subfig}
\usepackage{comment}
\usepackage{graphicx}%
\usepackage{dcolumn}%
\usepackage{bm}%
\usepackage[utf8]{inputenc}
\usepackage[T1]{fontenc}
\usepackage{mathptmx}
\usepackage{amsmath}
\usepackage[capitalise]{cleveref}
\usepackage{xcolor}
\usepackage{xprintlen}
\usepackage{soul}

\newcommand{\vk}{{\boldsymbol{k}}}                        	
\newcommand{\e}[1]{\mathrm{e}^{#1}}

\newcommand{\vecg}{{\boldsymbol{g}}}

\newcommand{\veck}{{\boldsymbol{k}}}

\newcommand{\vecsigma}{{\boldsymbol{\sigma}}}
\newcommand{\vecn}{{\boldsymbol{n}}}

\newcommand{\ii}{\mathrm{i}}

\def\eg{\textit{e.g. }}

\begin{document}
\title{Colloquium: Spin-orbit effects in superconducting hybrid structures}
\author{Morten Amundsen}
\affiliation{\mbox{Nordita, KTH Royal Institute of Technology and Stockholm University,}} 
\affiliation{\mbox{Hannes Alfv\'{e}ns v\"{a}g 12, SE-106 91 Stockholm, Sweden}}
\affiliation{\mbox{Center for Quantum Spintronics, Department of Physics, Norwegian University of Science \& Technology,}}
\affiliation{\mbox{NO-7491 Trondheim, Norway }}
\email[Corresponding author:]{morten.amundsen@ntnu.no}

\author{Jacob Linder}
\affiliation{\mbox{Center for Quantum Spintronics, Department of Physics, Norwegian University of Science \& Technology,}}
\affiliation{\mbox{NO-7491 Trondheim, Norway}}
\email[Corresponding author:]{jacob.linder@ntnu.no}
 
\author{Jason W. A. Robinson}
\affiliation{\mbox{Department of Materials Science  \& Metallurgy, University of Cambridge, 27 Charles Babbage Road,}}
\affiliation{\mbox{Cambridge CB3 0FS, United Kingdom}}
\email[Corresponding author:]{jjr33@cam.ac.uk}

\author{Igor \v{Z}uti\'c}
\affiliation{\mbox{Department of Physics, University at Buffalo, State University of New York, Buffalo, New York 14260, USA}}
\email[Corresponding author:]{zigor@buffalo.edu}

\author{Niladri Banerjee}
\affiliation{\mbox{Department of Physics, Blackett Laboratory, Imperial College London, London SW7 2AZ, United Kingdom}}
\email[Corresponding author:]{n.banerjee@imperial.ac.uk}

\begin{abstract}
Spin-orbit coupling (SOC) relates to the interaction between an electron’s motion and its spin, and is ubiquitous in solid-state systems. Although the effect of SOC in normal-state phenomena has been extensively studied, its role in superconducting hybrid structures and devices opens many unexplored questions. In conjunction with broken symmetries and material inhomogeneities within superconducting hybrid structures, SOC may have additional contributions, beyond its effects in homogenous materials. Remarkably, even with well-established magnetic or nonmagnetic materials and conventional $s$-wave spin-singlet superconductors, SOC leads to emergent phenomena including equal-spin triplet pairing and topological superconductivity (hosting Majorana states), a modified current-phase relationship in Josephson junctions, and nonreciprocal transport. SOC is also responsible for transforming quasiparticles in superconducting structures which enhances the spin Hall effect and changes spin dynamics. Taken together, SOC in superconducting hybrid structures and the potential for electric tuning of the SOC strength, creates fascinating possibilities to advance superconducting spintronic devices for energy-efficient computing, and enable topological fault-tolerant quantum computing. By providing a description of experimental techniques and theoretical methods to study SOC, this Colloquium describes the current understanding of resulting phenomena in superconducting structures and offers a framework to select and design a growing class of materials systems where SOC plays an important role. 
\end{abstract}

\maketitle
\tableofcontents

\section{Introduction}
As a relativistic effect, the motion of an electron in an electric field creates a magnetic field in its rest frame~\cite{Jackson:1998}. The resulting spin-orbit coupling (SOC) in solid-state systems can have different contributions. In addition to the coupling of electron spin with the average electric field from the periodic crystal potential, other SOC terms arise due to an applied or built-in electric field, for example due to broken inversion symmetry. One can also distinguish intrinsic, extrinsic, and synthetic SOC, due to electronic structure, impurities, and magnetic textures, respectively. With SOC, at a given wave vector, {\bf k}, the twofold spin degeneracy is removed resulting in a $k$-dependent Zeeman energy and an effective magnetic 
field~\cite{Winkler:2003,Zutic2004:RMP}. In superconducting heterostructures, the role of SOC can be even more striking by transforming the orbital and spin symmetry of the Cooper pairs, through which exotic states may emerge---even from simple $s$-wave spin-singlet superconductors.

For decades, SOC effects have been identified as crucial for many normal-state phenomena, such as spin-photon and spin-charge conversion~\cite{Meier:1984}, various topological states 
\cite{Armitage2018:RMP,Shen:2012}, the family of spin Hall effects \cite{Dyakonov1971:PL,Maekawa:2012}, magnetocrystalline anisotropy and noncolinear spin textures (including skyrmions and chiral domain walls) \cite{Tsymbal:2019}. They also formed the basis for early spintronic applications, which can be 
traced back to the discovery of anisotropic magnetoresistance in 1857 \cite{Thomson1857:PRSL,Zutic2004:RMP}. In contrast, the relevance of SOC in superconducting structures was largely absent, or limited to specific aspects without fully recognizing many 
connections~\cite{bergeret_rmp_05,Buzdin2005,golubov_rmp_04, Tedrow1971:PRL,Tedrow:1994}. Motivated by recent advances in studies of hybrid superconducting structures where SOC plays a prominent role, this review aims to provide an experimental and theoretical framework to highlight many such connections between different phenomena and emerging applications in these structures. 

The quest to realize topological superconductivity and elusive Majorana states for fault-tolerant topological quantum computing in structures with strong SOC relies on equal-spin-triplet superconductivity~\cite{Nayak2008b,Elliott2015:RMP}. This triplet superconductivity is also sought  in superconducting 
spintronics~\cite{Linder2015,Eschrig2015a,OhnishiBoost,YangBoost} 
as it supports dissipationless spin currents and allows for the coexistence of superconductivity and ferromagnetism. 
Josephson junctions (JJs) with tunable SOC, which enable spin-triplet superconductivity, are important building blocks for topological superconductivity and superconducting 
spintronics~\cite{Mayer2020:NC,Dartiailh2021:PRL}. These JJs also reveal the superconducting diode effect~\cite{Dartiailh2021:PRL,Baumgartner2022:NN} an example of a nonreciprocal phenomenon~\cite{Nadeem2023:P}. While nonreciprocal effects are technologically important~\cite{Marder:2010,Shockley1952:PT} and known since the nineteenth century in the normal state~\cite{Faraday1846:PTRS,Kerr1877:PM}, experimental demonstrations of superconducting counterparts were largely absent, until a few years ago~\cite{Nadeem2023:P}. 
Analogous to 
multiferroic materials which allow electrical control of magnetic properties and, conversely, magnetic control of electrical properties, we can view SOC in the superconducting state as enabling various magnetoelectric effects~\cite{Tkachov2017:PRL} and facilitating the coupling between different order parameters. 
Since 
SOC changes the properties of 
quasiparticles in superconductors, it has also been shown to produce strongly enhanced spin Hall phenomena in superconducting 
structures.

With controllable SOC, the previous efforts to integrate superconductors and ferromagnets can be radically simplified. Instead of engineering complex noncollinear magnetic structures at the superconductor/ferromagnet (S/F) interface~\cite{keizer_nature_06,Khaire2010,Robinson2010,Robinson:2012,Usman_2011_PRB,BanerjeeRev:NatComm}, a single common F with SOC in a superconducting heterostructure with broken inversion symmetry is sufficient to support spin-triplet superconductivity and large magnetoresistive 
effects~\cite{Banerjee2018,Martinez:2020,Cai2021:NC,Gonzalez:2021,Gonzalez:2020,Jeon2018,Jeon:2019,Jeon:2020}. Theoretically, the observed role of
SOC in singlet-to-triplet pair conversion has been studied for both ballistic and diffusive 
transport~\cite{feng_jap_08,Hogl2015,yokoyama_prb_06,bergeret_prl_13,bergeret_prb_14,Jacobsen2015} and
preceded by the related effect of spin-active interfaces~\cite{Halterman2009:PRB,Linder:2009,Zutic1999:PRBa,Eschrig2003:PRL} and SOC generated $k$-anisotropic 
triplet condensates~\cite{Edelstein:2003,gorkov_prl_01}.

Another example where SOC fundamentally modifies the underlying physics is within superconducting random-access memories using ferromagnetic JJs.
Here, nonvolatile control of the zero and $\pi$ ground state phase encoding binary information~\cite{Dayton2018:IEEEML,Birge2019:IEEEML}, needs to be revisited in the presence of SOC where, in addition to the spin-singlet and the spin-triplet states, their admixture is also possible. The resulting anomalous Josephson effect~\cite{buzdin_prl_08,Reynoso2008:PRL} supports an arbitrary phase shift other than just zero and $\pi$, leading to novel challenges and opportunities for non-binary information processing and storage. Just as magnetic JJs are the building blocks for superconducting memories, their nonmagnetic counterparts are the key elements for low-power and high-speed superconducting logic~\cite{Holmes2013:IEEE,Tafuri:2019} and superconducting quantum computing~\cite{Krantz2019:APR,Wendin2017}. This means that SOC may not only modify such devices, but also provide entirely new functionalities in their operation, as current-phase relation, Josephson energy, critical temperature and critical current, can all strongly change with SOC. As in the normal state, SOC is the major source of spin relaxation and decoherence, as well as the underlying mechanism for spin dynamics, in the superconducting state~\cite{Zutic2004:RMP}. Since both long and short spin relaxation times~\cite{Nishikawa1995:APL,Lindemann2019:N} can be desirable in the normal state, their SOC-controlled tunability in the superconducting state would be similarly advantageous. 
Taken together, the presence of SOC and its tunability in hybrid superconducting structures offers an intriguing prospect to both identify novel phenomena as well as advance various quantum technologies, from storing, transferring, and processing information, to improving quantum sensing. While some of the resulting efforts simply extend the current concepts and applications of superconductivity, others, like topological quantum computing, would radically change the paths towards realizing computational architectures~\cite{Cai2023:AQT}. Even if the most ambitious proposals remain aspirational, the advances in our understanding of SOC have already transformed the way we view various superconducting phenomena.

By focusing on more common materials, 
where their superconductivity is well established, simplifies the understanding of the role of SOC. For example, a large part of this review focuses on hybrids with conventional elemental or nitride $s$-wave superconductors, including  Nb, Al, V, and NbN. However, we  
note that other superconducting systems to investigate SOC effects are possible; for over two decades superconducting spintronics has been studied with high-temperature $d$-wave 
superconductors~\cite{Vasko1997:PRL,Wei1999:JAP,Chen2001:PRB}, where even their normal-state properties remain debated.  
 In briefly mentioning other oxide superconductors, two-dimensional superconductors, such as NbSe$_2$, and proximity-induced superconductivity in III-V and group IV semiconductor nanostructures, we complement our Colloquium by providing relevant reviews on these topics.

While we recognize that significant development is required before SOC-driven superconducting phenomena can be applied in the field of spintronics or quantum technologies, sufficient progress has already been made both theoretically and experimentally where one can start to think about potential areas of application. 
Using dissipationless supercurrents offers alternatives for energy-efficient information-communication and quantum technologies. Just data centers alone are predicted to require 8\% of globally generated electrical power by 2030~\cite{Jones2018:N}.  A potential solution may combine superconducting electronics with recent advances in 
spintronics~\cite{Hirohata2020,Tsymbal:2019} to seamlessly integrate logic and memory~\cite{Birge2019:IEEEML} and thereby overcome the von Neumann 
bottleneck~\cite{Dery2012:IEEETED}.
More importantly, through this review we hope to drive future innovations benefiting fields like superconducting spintronics and Majorana physics which will assist in solving material challenges that are key to progress in these areas. 

We start with a brief introduction to relevant theoretical and experimental 
background of superconductivity in the presence of SOC in hybrid structures, 
followed by recent developments and, finally, concluding with 
open questions and highlight promising research directions.

\section{Background}

This section reviews basic concepts which build a basis for the results outlined in the following sections. We start by describing SOC, specifically the Rashba and Dresselhaus models, in bulk materials and structures with inversion asymmetry. We then introduce the physics emerging from SOC superconductivity in proximity structures and include a discussion on theoretical 
methods that can be used to study such systems. The purpose of this section is thus to provide the reader with a set of theoretical concepts necessary to discuss the interesting spintronics phenomena that arise due to the combination of superconductivity and SOC in heterostructures.

\subsection{Spin--orbit coupling}
\label{SOC}
Coupling between the motion of an electron and its spin stems from the fact that in the reference frame of the electron, it is the positively charged lattice that moves. Moving charges create a magnetic field, which may couple to the electron spin. In a Lorentz-invariant formulation, a SOC term emerges, as shown in the Dirac equation~\cite{dirac_proc_28}
\begin{align}
\begin{pmatrix} mc^2 + V(\bm{r}) & -i\hbar c\bm{\sigma}\cdot\nabla \\ -i\hbar c \bm{\sigma}\cdot\nabla & -mc^2 + V(\bm{r}) \end{pmatrix} \begin{pmatrix} \psi_e \\ \psi_h\end{pmatrix} 
= \left(\varepsilon + mc^2\right)\begin{pmatrix} \psi_e \\ \psi_h\end{pmatrix},
\end{align}
and taking the nonrelativistic limit, $\varepsilon,V\ll mc^2$, where $\varepsilon$ is the particle energy without its rest mass.  
Here, $V(\bm{r})$ is the lattice potential, 
$m$ is the free electron mass, and $\boldsymbol{\sigma}$ is a vector of Pauli matrices. 
The resulting Hamiltonian is
$H = p^2/2m + V(\bm{r}) + \hbar/(4m^2c^2)\bm{\sigma}\cdot\left(\nabla V \times \bm{p}\right),$ for the electron wavefunction $\psi_e$, where irrelevant terms are discarded. 
The last term represents SOC, 
large near a lattice site 
~\cite{Fabian2007:APS}.
Within second quantization, in the basis of the Bloch functions, it takes the form~\cite{Samokhin2009},
\begin{align}
H_{SO} = \sum_k\sum_{nn'}\sum_{ss'} \bm{Q}_{nn'}(\bm{k})\cdot\bm{\sigma}_{ss'}c_{kns}^\dagger c_{kn's'},
\label{eq:hso}
\end{align}
where $\bm{Q}_{nn'}$ is a phenomenological model function which expresses the coupling between momentum and spin, with $n$ and $n'$ band indices, and $\bm{k}$ the crystal momentum. 
For a centrosymmetric material,
the terms $\bm{Q}_{nn}$ vanish, 
SOC can only be described by models containing at least two bands. However, in a noncentrosymmetric material a one-band model is possible, $\sum_{nn'} \bm{Q}_{nn'}\to \bm{Q}$ in \cref{eq:hso}, while 
$\bm{Q}(\bm{k}) = - \bm{Q}(-\bm{k})$.
One can distinguish bulk and structure inversion asymmetry (BIA, SIA), 
which lead to a spin splitting and SOC
\begin{align}
H_{SO}(\bm{k}) = \hbar \bm{\sigma} \cdot \bm{\Omega}(\bm{k})/2, 
\label{eq:Larmor}
\end{align}
where $\bm{\Omega}(\bm{k})$ is the Larmor frequency for the electron spin precession in the conduction band~\cite{Zutic2004:RMP} or, equivalently, SOC field. 
Here momentum scattering, 
$\bm{\Omega}(\bm{k})$ is responsible for spin dephasing.
Related SOC manifestations in semiconductors usually focuses on effective models which capture the
low-energy properties of the conduction and valence bands. An 
example of BIA is the Dresselhaus SOC~\cite{Dresselhaus1955},
given by $\bm{\Omega}_D=(2\gamma/\hbar)[k_x(k_y^2-k_z^2), k_y(k_z^2-k_x^2),k_z(k_x^2-k_y^2)],$
where $\gamma$ is the SOC strength. In two-dimensional (2D) systems with quantum
confinement along the unit vector $\hat{n}$, $\bm{\Omega}_D$ can be linearized in ${\bm k}$, 
\begin{align}
\bm{\Omega}^\text{2D}_D \sim k_n^2[2 n_x(n_y k_y-n_z k_z)+k_x(n_y^2-n_z^2)]{\bf{\hat x}}+ c.p.,
\label{eq:2D}
\end{align}
where $k_n^2$ is the expectation value of the square of the 
wave number operator normal to the plane in the lowest subband state while $\hat{\boldsymbol{n}} = (n_x,n_y,n_z)$ is the confinement unit vector of the quantum well,
and c.p. denotes the cyclic index permutation. For a rectangular well of 
width $a$, $k_n^2=(\pi/a)^2$, 
while for 
a triangular well 
$k_n^2$ is given in \onlinecite{deSousa2003:PRB}. 
With a strong confinement, $k_\|^2\ll k_n^2$, where $\bm{k}_\|$
is the in-plane (IP) wave vector ($\perp \hat{n}$),
cubic terms in ${\bm k}$ in $\bm{\Omega}_D$ from Eq.~(\ref{eq:2D})
can be neglected. 

\begin{figure}[t]
    \centering
    \includegraphics[width=0.8\columnwidth]{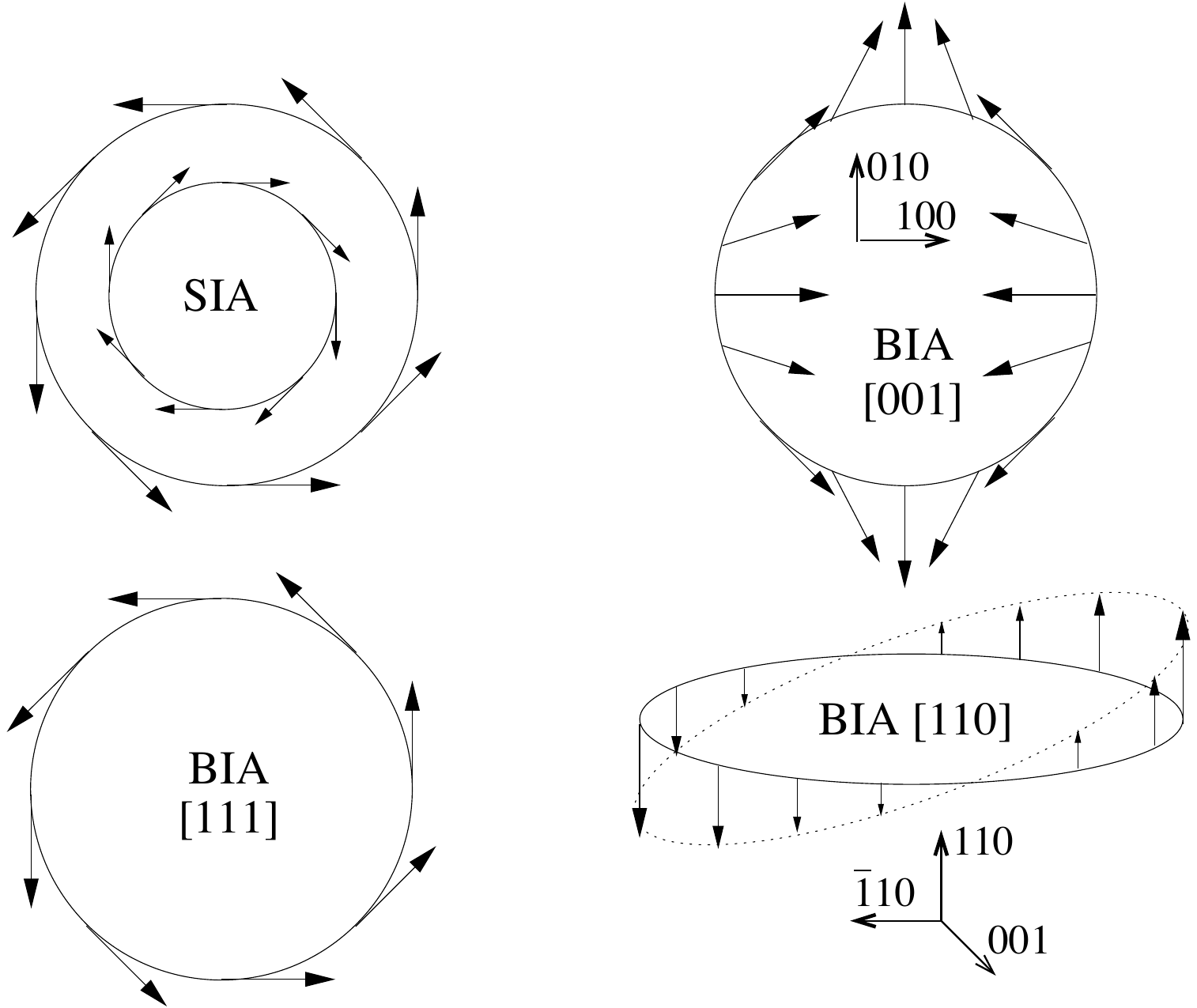}
    \caption{Vector fields ${\bm\Omega}({\bm k})$ on a circular Fermi surface 
    for structure (SIA) and bulk (BIA) inversion asymmetry. Since 
    ${\bm\Omega}({\bm k})$ is the spin quantization axis, the vector pattern is also the pattern of the spin on the Fermi surface. As opposite spins have different energies, the Fermi circle splits into two concentric circles with opposite signs of spin; shown here only for the SIA case, but the analogy extends to all examples. The field for BIA [110] perpendicular to the plane, with the magnitude varying along the Fermi surface. All other cases have constant fields lying in the plane. From~\onlinecite{Zutic2004:RMP}.}
    \label{fig:omega}
\end{figure}

For commonly considered quantum well confinements, one obtains for [001]: $\bm{\Omega}^\text{2D}_D\sim k_n^2(-k_x,k_y,0)$, 
for [111]: $\bm{\Omega}^\text{2D}_D\sim k_n^2(\bm{k}\times \bm{n})$, and for
[110]: $\bm{\Omega}^\text{2D}_D\sim k_n^2 k_x(-1,1,0)$, as shown in Fig.~\ref{fig:omega}.
Several features can be readily seen, BIA [100] displays a ``breathing" pattern, while BIA [110] ${\bm\Omega}({\bm k})$ is perpendicular to the plane such that, within the linear in ${\bm k}$ approximation, the perpendicular spins do not dephase. 

An extensively studied SIA example is given by Bychkov-Rashba (or just Rashba) SOC~\cite{Bychkov1984}, which arises in asymmetric quantum wells or in deformed bulk systems, expressed by
\begin{align}
\bm{\Omega}_R=2 \alpha (\bm{k}\times \bm{n}),
\label{eq:Rashba}
\end{align}
where $\alpha$ parametrizes its strength and the inversion
symmetry is broken along the ${\bm n}$-direction. We see in Fig.~\ref{fig:omega} that
its functional form, $\bm{\Omega}_R$, coincides with BIA $\bm{\Omega}^\text{2D}_D$ in
[111] quantum wells. A desirable property of Rashba SOC is that
$\alpha$ is tunable by an applied electric field.
While these linearized forms of the Rashba and Dresselhaus SOC are the most common models, there is an increasing interest in the study of phenomena which go beyond their range of validity.

For example, in Rashba SOC there is a growing class of materials where cubic terms in $k$ can play an important role, or even be dominant~\cite{Alidoust2021:PRB}. Finally, while we have here focused on intrinsic SOC, we note that extrinsic SOC caused by impurities plays a key role in spintronics both with and without superconductors, giving rise to important contributions to spin Hall effects and spin relaxation~\cite{Zutic2004:RMP}.

\begin{figure}
\resizebox{17cm}{!}{\includegraphics{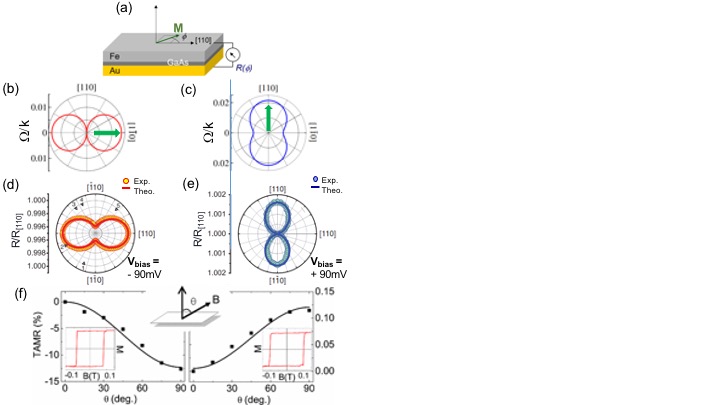}} 
\caption{(a) Schematic 
of a Fe/GaAs slab. For in-plane TAMR, $\mathbf{M}$, is rotated in the plane of Fe. (b) Angular $\mathbf{k}$-space dependence of the amplitude of the interfacial SOC field for {\bf M} along the GaAs $[1\bar{1}0]$ direction (green arrow). (c) Same as in (b) but for $\mathbf{M}$ along the [110] direction~\cite{Gmitra2013:PRL}. The 
tunneling resistance $R(\phi)$ is normalized to its $\phi = 0$ value, $R_{[110]}$. Measurements for bias $\pm 90$~meV are shown in (d) and (e), respectively~\cite{Moser2007:PRL}. (f) Angular dependence of the TAMR in the out-of-plane (OOP) configuration. Left (right) panels correspond to CoPt/AlO$_x$/Pt (Co/AlO$_x$/Pt) tunnel junctions. An extra Pt layer with strong SOC yields in CoPt/AlO$_x$/Pt two orders of magnitude larger TAMR than in Co/AlO$_x$/Pt. The insets show {\bf M} measurements in OOP magnetic fields~\cite{Park2008:PRL}. 
Adapted with permission from~\onlinecite{Zutic2019:MT}.}
\label{fig:ISOC}
\end{figure}

\subsection{Validity of Rashba and Dresselhaus models}

The validity of effective low-energy SIA and BIA
SOC models
can be examined from electronic structure calculations, using first-principles, ${\bm k\cdot \bm p}$ method, or a tight-biding model. 
Another contribution to Rashba-like spin splitting, $H_L =-\bf{p}\cdot \bf{E}$, 
arises in systems with localized orbital momentum, 
$\bf{L}$~\cite{Park2011:PRL,Park2012:PRB}, 
where $\bf{E}$ is the electric field and the electric dipole moment, 
$\bf{p} \propto \bf{L} \times \bf{k}$,
is produced by the asymmetric charge distribution. 
Rashba SOC strength is 
renormalized by the orbital contribution, while local symmetry breaking can induce local
orbital Rashba spin splitting even in centrosymmetric systems~\cite{Lee2020:npj2DMA}. 

Going beyond Rashba and 
Dresselhaus model might be necessary for interfacial SOC in junctions with 
interface-induced
symmetry reduction in the individual bulk constituents and for multiorbital configurations \cite{mercaldo_prapp_20}. This is exemplified in an Fe/GaAs junction,
where the cubic and $T_d$ symmetries of bulk Fe and GaAs, respectively, 
are reduced to $C_{2v}$
~\cite{Fabian2007:APS,Zutic2019:MT}.  

Since the interfacial SOC is present only near the interface, its effects can be controlled electrically via a gate voltage or an applied external bias capable of pushing the carrier wave function into or away from the interface. Interfacial SOC can also be controlled magnetically, as it strongly depends on the 
the orientation of {\bf M} in the Fe layer, as illustrated from the first-principles calculation in Figs.~\ref{fig:ISOC}(b) and (c)~\cite{Gmitra2013:PRL}. 
The bias-dependence of the SOC can be inferred from the
transport anisotropy in Figs.~\ref{fig:ISOC}(d) and (e).

While the resulting interfacial SOC for Fe/GaAs junction corresponds neither to Rashba, nor Dresselhaus
models, its existence can be probed through 
tunneling anisotropic 
magnetoresistance (TAMR), which gives the dependence of the tunneling current in a junction with only \emph{one} magnetic electrode on the orientation of 
$\mathbf{M}$~\cite{Gould2004:PRL}. 
For an IP rotation of $\mathbf{M}$ in Fig.~\ref{fig:ISOC}(a), 
we can define TAMR as the normalized resistance difference, 
\begin{equation}
\mathrm{TAMR}=(R(\phi)-R_{[110]})/R_{[110]},
\label{eq:TAMR}
\end{equation}
where $R(\phi=0)\equiv R_{[110]}$ is the resistance along the [110] crystallographic axis. The out-of-plane (OOP) TAMR
is defined analogously.
TAMR appears because the electronic structure depends on the $\mathbf{M}$ orientation, due to SOC. The  surface or an interface electronic structure can strongly deviate from its bulk counterparts and host pure or resonant bands. 
With SOC, the dispersion of these states depends on the $\mathbf{M}$ orientation~\cite{Chantis2007:PRL}. As a result, the tunneling conductance, which, in a crystalline junction, is very sensitive to the transverse wave vector, develops both OOP and IP MR, shown in Figs.~\ref{fig:ISOC}(d)-(f),
whose angular dependence reflects the crystallographic symmetry of the interface. For example, the TAMR inherits the $C_{4v}$ symmetry for the Fe (001) surface~\cite{Chantis2007:PRL} and the reduced $C_{2v}$ symmetry for the Fe(001)/GaAs interface~\cite{Moser2007:PRL}. 

Our prior discussion of SOC and its manifestations in the normal-state properties have important superconducting
counterparts as well as enable entirely new phenomena, absent in the normal state. Even when the SOC results in only a small normal-state transport anisotropy, as shown in Figs.~\ref{fig:ISOC}(d) and (e), the superconducting analogs of such phenomena can lead to much greater effects~\cite{Cai2021:NC,Martinez:2020}.

\subsection{Triplet superconductivity}
Conventional $s$-wave superconductors are well-described by the Bardeen-Cooper-Schrieffer (BCS) microscopic theory. The superconducting correlations consist of Cooper pairs in a spin-singlet state and 
carry no net spin, unlike the proximity-induced spin-triplet superconducting correlations. 
There are materials believed to exhibit intrinsic triplet superconductivity such as 
Bechgaard salts~\cite{Sengupta2001:PRB}, UPt$_3$~\cite{Joynt2002}, as well as the ferromagnetic superconductors~\cite{Aoki2011}. A direct interaction between superconductivity and SOC is also found in noncentrosymmetric superconductors, where electron pairing is a mixure of spin-singlet and spin-triplet ~\cite{Smidman2017}.
 
Through proximity effects, triplet superconducting correlations can
be generated using only conventional materials. In S/F bilayers, the spin splitting in the latter leads to oscillations in the pair correlation between the singlet and triplet spin configurations due to a process akin to the Fulde-Ferrell-Larkin-Ovchinnikov (FFLO) oscillations~\cite{Fulde1964,Larkin1965,Buzdin2005}. Nevertheless, such a coupling between S and homogeneous F is rapidly suppressed as one moves away from the interface region, leading to a short-range proximity effect. The situation is  different in F with an inhomogeneous {\bf M} direction, where the spin of the short-ranged triplet correlations is orthogonal to {\bf M}. Here, the short-ranged triplets decay over the coherence length of the Cooper pairs in the F layer. If the orientation of {\bf M} changes, the triplet spin will obtain a component parallel to {\bf M}. This component, referred to as a long-ranged triplet component, is not influenced by the spin splitting to the same degree as their short-ranged counterparts. Remarkably, it may persist for long distances as in nonmagnetic metals~\cite{PhysRevB.58.R11872,Petrashov1994:JETP,Lawrence_1999,Petrashov_1999_PRL},  of the order $\sqrt{D/{2\pi T}}$ in the diffusive limit, where $D$ is the diffusion coefficient of the F region and $T$ is the temperature~\cite{bergeret_rmp_05,bergeret_prl_01,kadigrobov_epl_01}. 

Engineering superconducting hybrid structures for generating spin-polarized triplets have been extensively studied. Using magnets with rare earth materials (holmium) with intrinsically inhomogeneous {\bf M} in JJs, provides evidence of triplet pair creation~\cite{Sosnin2006,Robinson2010}. Alternatively, noncollinear magnetism can be engineered in magnetic multilayers~\cite{Khaire2010,BanerjeeRev:NatComm}. However, intrinsically inhomogeneous magnets are rare and controlling  {\bf M}-misalignment in ferromagnetic multilayers is difficult.
Magnetic vortices are emerging as a viable candidate for tunable sources of  noncollinear magnetism and corresponding triplet generation~\cite{Kaveh_2022_vortex,PhysRevResearch.4.033136}.

An effective inhomogeneous {\bf M} is generated in a homogeneous F in the presence of SOC. This takes the form of a Dzyaloshinskii-Moriya exchange interaction~\cite{Dzyaloshinsky1958,Moriya1960}, which cants {\bf M} creating helical spin textures~\cite{Ferriani2008} and skyrmions~\cite{Rossler2006,Heinze2011}. Such magnetic structures at S/F interface can generate spin-polarized triplets. Furthermore, 
the formation of triplet in S/F structures with an interfacial SOC generated
due to broken inversion symmetry is an area of intense study ~\cite{MelNikov2012,bergeret_prl_13,bergeret_prb_14,Jacobsen2016,Jeon2018,Banerjee2018,satchell_prb_18, simensen_prb_18, satchell_prb_19,Jeon:2019,Jeon:2020,Hogl2015,Vezin2020,Cai2021:NC}.

\subsection{Theoretical frameworks}
In the following we describe common theoretical models which describe SOC in superconducting hybrid structures with 
increasing realism. We start by highlighting basic features of the proximity effect in such systems.

\subsubsection{General considerations of spin-dependent fields}
\label{sec:SCh}
\begin{figure}[b]
    \centering
    \includegraphics{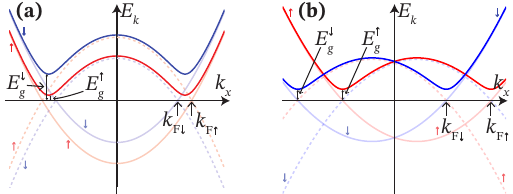}
    \caption{The band structure of a superconductor with 
    (a) a Zeeman field, (b) SOC of the form $H_{SO} = \alpha k_x\sigma_z$. $E_g^s$ indicates the superconducting gap of the two branches. The faded, full (dashed) lines: 
    the normal-state electron (hole) band structures of the two spins. The arrows accompanying each band indicate its spin index.} 
    \label{fig:proxf}
\end{figure}

To understand the response of a superconductor to magnetic interactions, we consider the BCS model in an infinite domain, with spin splitting of the type
\begin{align}
H &= \sum_{ks} \xi_k c_{ks}^\dagger c_{ks} - \sum_{ks} \left[s\Delta c_{ks}^\dagger c_{-k,-s}^\dagger + s\Delta^*c_{-k,-s}c_{ks}\right] \nonumber \\
&- \sum_{kss'} \bm{h}(\bm{k})\cdot\bm{\sigma}_{ss'}c_{ks}^\dagger c_{ks'},
\end{align} 
where $\xi_k = \hbar^2\bm{k}^2/2m - \mu$, $\mu$ is the chemical potential, and $m$ is the electron rest mass. $\Delta$ is the superconducting order parameter, and $\bm{h}$ relates to the spin splitting. The operators $c_ {ks}^\dag\; (c_{ks})$ create (annihilate) an electron with momentum $k$ and spin $s$. Consider the case where $\bm{h} = h_0\hat{z}$ is independent of momentum, and therefore reduces to a Zeeman field. Insight into the superconducting behavior can be gained by inspecting the normal state dispersions ($\Delta = 0$) as shown in \cref{fig:proxf}(a). The Zeeman field splits the energy bands of the two spin species---the spin-up band is lowered in energy, and the spin-down band raised. The corresponding gap induced in the band structure of the two spin species when superconductivity is introduced, $E_g^{\uparrow}$ and $E_g^{\downarrow}$ respectively, is $E_g^{s} = \Delta - sh$, and indicates a difference in the strength of the hybridization between electrons and holes for the two opposite spins. We can study the effect of superconductivity via the opposite-spin electron pair correlations, given as $F_{\uparrow\downarrow}(\bm{k},t) = \left\langle c^\dagger_{k\uparrow}(t)c^\dagger_{-k\downarrow}(0)\right\rangle$. We note that this quantity can represent a scattering process between a spin-down hole and a spin-up electron, and therefore involves no exchange of spin. This is because $c^\dagger_{-k\downarrow}$ creates a spin-down electron, which is equivalent to the removal of a spin-up hole. 

A similar analysis applies to $F_{\downarrow\uparrow}(\bm{k},t) = \left\langle c^\dagger_{k\downarrow}(t)c^\dagger_{-k\uparrow}(0)\right\rangle$, which now involves spin-down particles. The Zeeman field changes the relative size of the pair correlations. 
The superconducting correlations have to be antisymmetric under the combined interchange of spin, momentum, band, and time indices, a relationship known as the SPOT rule~\cite{Berezinskii1964,linder_rmp_19}. Singlet pairing (odd in spin index), $F_s\propto F_{\uparrow\downarrow} - F_{\downarrow\uparrow}$, is the most conventional form of superconductivity, and is typically $s$ wave (even parity), single-band (even in band index), and even frequency (even in time index). However, since $F_{\uparrow\downarrow}$ and $F_{\downarrow\uparrow}$ now are different, one obtains a triplet component $F_t \propto F_{\uparrow\downarrow} + F_{\downarrow\uparrow}$. Since we have no $k$ dependence in either the order parameter or the Zeeman field, we typically get $s$-wave correlations, meaning that the triplets must be odd frequency. Hence we derive $F_{s\bar{s}} = -s \Delta/[(i\omega +sh)^2 - \xi_k^2 -|\Delta|^2],$ with $\bar{s} = -s$,
from which we see that $F_t\propto ih\omega$, and odd when $\omega\to-\omega$. 
 
In the model considered here, even frequency triplets require oppositely aligned spins with mismatched momenta so that an odd-parity component appears. This is possible in the presence of a Zeeman field and a spatially modulated superconducting order parameter. This is the 
FFLO phase~\cite{Fulde1964,Larkin1965}, which has not been observed experimentally in the bulk. Such triplets are more readily obtained by considering a spin splitting of the form $\bm{h} = h_0k_x\hat{z}$, i.e., a form of SOC. The normal-state dispersions are shown in \cref{fig:proxf}(b), plotted along $k_x$. There is a relative horizontal shift of the energy bands of the two spins meaning that spin-up particles on average have a positive momentum along $k_x$, while spin-down particles have a negative momentum. Hence, even in the absence of {\bf M}, there exists an equilibrium spin current 
-- including in the normal state~\cite{Rashba2003,Tokatly2008:PRL,Sonin2007:PRB,Sonin2010:AiP, Droghetti2022}. The momentum shift of the normal-state dispersions leads to a similar, relative momentum-shift in the pair correlations $F_{s\bar{s}} = -s\Delta/[(i\omega)^2 - (\xi_k-sh_0k_x)^2 -|\Delta|^2],$ which gives rise to $p_x$ wave triplets, $F_t\propto h_0k_x$.   

When $F_{\uparrow\downarrow}\neq F_{\downarrow\uparrow}$ it means that one spin species has a greater hybridization with their corresponding hole branch than the other. In other words, we have spin-dependent scattering processes in the electron-hole sector, and hence might expect an observable {\bf M} to appear. However, to obtain {\bf M}, it is a requirement that the relative phase difference between the singlet and the triplet is \emph{different from} $\pi/2$, otherwise $F_{\uparrow\downarrow} \propto |F_s| + i|F_t|$ and $F_{\downarrow\uparrow} \propto |F_s| - i|F_t|$ remain equal in magnitude~\cite{linder_prb_17}. For the Zeeman field, the odd-frequency triplets indeed incur such a phase shift of $\pi/2$, and therefore do not directly contribute to {\bf M}. On the other hand, the even frequency triplets are not phase shifted relative to the singlet correlations. This produces a superconducting contribution to the spin currents, since $F_{\uparrow\downarrow} \propto |F_s| + |F_t|$ for $k_x > 0$, and $F_{\uparrow\downarrow} \propto |F_s| - |F_t|$ for $k_x < 0$, and vice versa for $F_{\downarrow\uparrow}$. In other words, there is preferential particle-hole scattering of one spin species for $k_x > 0$, and for the other spin species for $k_x < 0$.

\subsubsection{Superconducting proximity effect}

The superconducting proximity effect in a metallic material is enabled by the process of Andreev reflection~\cite{Buzdin2005,Zutic2004:RMP,Deutscher2005:RMP,Eschrig2018:PTRSA}. In this process, an incoming electron from the metallic side enters the superconducting material, crosses quasiparticle branch when it has penetrated far enough so that its energy equals the local value of the superconducting gap, and then travels back as a hole-like excitation which enters the normal metal. In the process of the incident electron crossing branch, a total charge of $-2e$ is transferred to the superconducting condensate, resulting in the creation of a Cooper pair. On the normal-metal side, the incoming electron becomes correlated to the hole that tunnels back into the normal metal, creating a superconducting phase-coherence that extends a long distance. 

\begin{figure}%
    \centering
\subfloat{{\includegraphics[width=3.5cm]{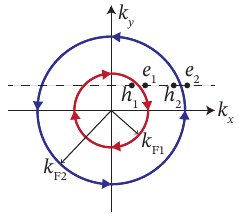} }}%
    \qquad
\subfloat{{\includegraphics[width=4cm]{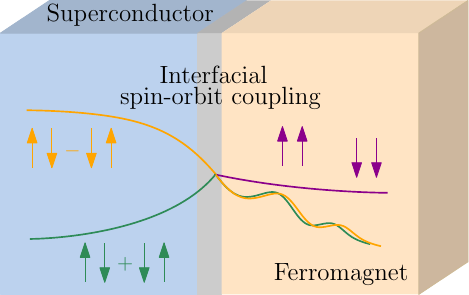} }}%
    \caption{\textit{Left}: Rashba SOC in 2D. The blue (outer) and red (inner) bands have opposite spins for a given angle in the $(k_x\,,\,k_y)$ plane. For an electron incident toward the interface with a direction away from the interface normal $(k_y\neq 0)$, Andreev 
    reflection is possible both via inter- and intraband scattering. Thus, the superconducting proximity becomes a mix of oscillatory and nonoscillatory terms inside the material with SOC. \textit{Right:} By adding a heavy metal (HM) layer with interfacial Rashba at the S/F interface, the proximity effect for spins parallel to {\bf M} can be extended through the generation of $\uparrow\uparrow$ and $\downarrow\downarrow$ triplet pairs. Conversely, for spins perpendicular to {\bf M} in F the spin-zero singlet and spin-zero triplet pairs remain short-ranged and oscillatory in F. The HM layer provides strong atomic SOC and SIA, 
    which produces an effective Rashba SOC 
    localized at the interface.}
    \label{fig:sfprox}%
\end{figure}

Before discussing how SOC modifies the proximity effect, consider 
F with the two bands split by an exchange field, modelled by 
$H= \boldsymbol{h} \cdot\boldsymbol{\sigma}$, where $\boldsymbol{h}$ is the exchange field. This causes the superconducting proximity effect to behave qualitatively differently compared to a normal metal~\cite{kadigrobov_epl_01, bergeret_prl_01, Buzdin2005, bergeret_rmp_05}. First of all, there will appear odd-frequency triplets due to the presence of superconducting correlations in a spin-split material. In addition, since the S/F interface breaks translation invariance, momentum in the direction normal to the interface is not a good quantum number. This leads to mixing between odd and even parity pair correlations, and thus odd parity, even frequency triplets. Consider the wave vectors of an electron excitation with a given spin, such as spin up, and a hole excitation with opposite spin at a given energy $\varepsilon$. The electron (hole) will have wave vectors $\pm k_\uparrow$ ($\mp k_\downarrow$). The mismatch $\Delta k = k_\uparrow - k_\downarrow$, between the electron and hole, gives the Cooper pair wavefunction induced in F
a finite center-of-mass momentum even in the absence of any net current through the system. Because of this, the superconducting correlations will not only decay as one moves deeper into F region, but they will also oscillate~\cite{Buzdin2005, bergeret_rmp_05}. This is shown in Fig.~\ref{fig:sfprox}. The decay is caused by the mismatch in wave vectors for the electrons and holes, causing a decoherence during propagation which reduces their correlation away from the interface.

We now discuss the superconductor proximity effect in hybrid structures with SOC~\cite{reeg_prb_15}. An important point which distinguishes the magnetic and SOC case is that the dimensionality is important for the qualitative behavior of the proximity effect. Consider first the one-dimensional (1D) case, \textit{e.g.} S/nanowire with SOC. 
We can model an antisymmetric Rashba-like SOC  
by $H_{SO} = \alpha \vecg_\vk \cdot \vecsigma,$
where $\alpha$ is the magnitude of the SOC and $\vecg_\vk = (0,0,k_x)$ for a nanowire extending along the $x$-axis \cite{cayao_prb_18}. We again consider an electron and hole with opposite spins at an energy $\varepsilon$ in \cref{fig:proxf}(b), as these are the excitations involved in the Andreev-reflection process inducing superconductivity in the SOC metal. The band-structure in the SOC metal is now different than in the ferromagnetic case. By considering electrons and holes with opposite spin labels, we see that such pairs have momenta $\pm k_{F\uparrow}$ or $\pm k_{F\downarrow}$, respectively. Therefore, there is no mismatch\footnote{When the Andreev reflection process involves quasiparticles with excitation energy $\varepsilon$, there is a tiny mismatch between the wavevectors due to different signs for $\varepsilon$ in the wavevector expression for electrons and holes. This mismatch also occurs in the normal metal case, and does not cause oscillations of the energy-integrated superconducting correlation function in the normal metal.} in the momentum magnitude between the electrons and the Cooper pair wavefunction does not acquire any center-of-mass momentum. Thus, Andreev reflection only involves intraband (same band) excitations in the 1D case and there is no oscillatory behavior of the superconducting correlations inside the SOC metal. In the diffusive limit, a decay length $\hbar/m\alpha$ of the triplet correlations in a SOC metal~\cite{reeg_prb_15} 
can be compared to the decay length $\sqrt{D/h}$ of triplets in F.

The situation changes qualitatively when going to two-dimensional (2D). We can model a 2D system with Rashba SOC by the same $H_{SO}$, but this time with a $\vecg_\veck$-vector that depends on both $k_x$ and $k_y$ for a system that lies in the $xy$-plane, such as a 2DEG. A Rashba-like SOC is described by $\vecg_\vk = (-k_y,k_x,0)$ and gives rise to the band-structure shown in \cref{fig:sfprox}.

As shown in the figure, the Fermi surface consists of two circles where the spin expectation value of an excitation on one of the circles varies as one moves around the circle. 
Assume for simplicity that we are dealing with a ballistic SOC/S structure so that translational invariance is maintained in the direction parallell to the interface. Then, momentum in this direction is conserved during the Andreev reflection process (say $k_y$ for a structure extending along the $x$-axis). Considering first the case $k_y=0$, we recover the 1D situation shown in \cref{fig:proxf}(b). But for $k_y\neq 0$ (black dashed line in left panel of \cref{fig:sfprox}, the proximity effect changes its nature. Since the Fermi surfaces do not have definite spin, an electron on the outer, blue circle $(e_2)$ can be Andreev reflected as a hole both on the outer blue $(h_2)$ and inner red ($h_1$) Fermi surface. Both of these holes carry some weight of opposite spin to the $e_2$ electron when $k_y\neq 0$, whereas only a hole on the same Fermi surfaces has opposite spin when $k_y=0$. As a result, both intra- and interband Andreev scattering
are possible. The intraband scattering gives rise to a nonoscillatory superconducting correlation decaying inside the SOC metal, like in the 1D case. But the interband Andreev scattering is seen to feature a momentum magnitude mismatch between the electron and holes involved: $k_{F1} - k_{F2}$. Thus, for interband Andreev scattering we are back to a similar situation as in the ferromagnetic case, where the induced superconducting correlations oscillate. 
The superconducting proximity effect in a SOC metal consists of both oscillatory and nonoscillatory terms, in contrast to both the ferromagnetic and 1D SOC metal case. 
Odd-frequency triplet superconductivity due to SOC has also been studied in topological insulators \cite{cayao_prb_17, cayao_prb_22}.\\ 

In Cooper pairs consisting of electrons with spins that are collinear with {\bf M} in F 
penetrate a long distance. This is because there is no longer any momentum mismatch between such electrons, as both belong to the same spin-polarized Fermi surface. 
This long-ranged superconducting proximity effect in F can be achieved by incorporating SOC e.g., (1) adding an interfacial heavy-metal (HM) layer causing SOC-scattering, or (2) F with intrinsic SOC. Case (1) is illustrated in Fig. \ref{fig:sfprox} where spin-dependent scattering at the interface due to the HM layer creates long range triplet pairs in F. Such pairs can survive $\sim 1\;\mu$m, even in strongly-polarized F~\cite{keizer_nature_06}. The number density of triplet pairs created in this way will depend on the direction of {\bf M} in the F layer~\cite{Jacobsen2015}. In case (2), long-ranged triplet pairs are also created, via two physical mechanisms~\cite{bergeret_prl_13}: spin-precession induced by the SOC, and anisotropic spin-relaxation. One can map the diffusive-limit equation of motion for the anomalous Green functions (describing the Cooper pairs), known as the Usadel equation, to the spin-diffusion equation in~\cite{bergeret_prb_14}. This analogy is useful since it shows that the different triplet Cooper pair components behave similarly to the spin components of an electron in a diffusive metal with 
SOC. Finally, there exists an interplay between the spin and orbital degrees of freedom even in the absence of SOC when a F couples to an intrinsic triplet S \cite{gentile_prl_13}.

\subsubsection{The Ginzburg-Landau formalism}\label{sec:gl}

The Ginzburg-Landau formalism is a symmetry-based method to explore the behavior of  superconducting systems~\cite{Ginzburg1950}. It involves expanding the free energy in the complex superconducting order parameter $\psi$, indicating the strength of the superconductivity, and is valid close to the superconducting transition temperature ($T_c$). The method is highly successful, and consistent with BCS theory~\cite{Gorkov1959}. Since $\psi$ describes singlet superconductivity, this approach cannot be used to obtain information about the triplet correlations, and thus is of limited use in making predictions relevant for superconducting spintronics, such as the generation of long-ranged triplets. Nevertheless, it is still possible to indirectly extract some information about spin-dependent phenomena---through their influence on $\psi$. A prominent example of this is the inclusion of a Rashba-type SOC and an exchange field $\bm{h}$, in which case 
the free energy density is given as~\cite{Edelstein1996:JPCM,Samokhin2004,Kaur2005}

\begin{align}
f(\bm{r}) =& a |\psi(\bm{r})|^2 + \gamma|\tilde{\nabla}\psi(\bm{r})|^2 + \frac{b}{2}|\psi(\bm{r})|^4 + \frac{\bm{B}^2}{2\mu_0} \nonumber \\
-&i\alpha(\bm{r})\left(\bm{n}\times\bm{h}\right)\left[\psi^*(\bm{r})\tilde{\nabla}\psi(\bm{r}) - \psi(\bm{r})\left(\tilde{\nabla}\psi(\bm{r})\right)^*\right],
\label{eq:GL}
\end{align}
where $\tilde{\nabla} = \nabla - (2ie/\hbar)\bm{A}$, and $\bm{B} = \nabla\times\bm{A}$. Furthermore $a$, $b$, and $\gamma$ are phenomenological parameters, $\vecn$ is 
unit vector along the direction of broken inversion symmetry, and $\alpha$ characterizes the SOC magnitude (in general, it can have a non-linear dependence on the atomic SOC strength). 
Applying \cref{eq:GL} to a Josephson weak link, a non-zero phase difference appears between the superconducting banks~\cite{buzdin_prl_08}. This is seen by minimizing \cref{eq:GL} with respect to $\psi$ and $\bm{A}$, giving the Euler-Lagrange equation
\begin{align}
&a\psi - \gamma\tilde{\nabla}^2\psi + b\psi|\psi|^2 - 2i\alpha\left(\bm{n}\times\bm{h}\right)\cdot\tilde{\nabla}\psi = 0, \label{eq:GLEL1}\\
&\bm{j} = \frac{4e\gamma}{\hbar}\Im\left(\psi^*\nabla\psi\right) - \left(\frac{8e^2\gamma}{\hbar^2}\bm{A} + \frac{4e\alpha}{\hbar}\left(\bm{n}\times\bm{h}\right)\right)|\psi|^2.
\label{eq:GLEL2}
\end{align}
To derive the above equation for $\psi$ microscopically, one may derive the quasiclassical Eilenberger equation (to be discussed below), find the solution for the anomalous Green function perturbatively in orders of $\Delta$, and then insert this solution into the self-consistent gap equation. Equation~(\ref{eq:GLEL1}) 
may be solved in the normal metal, where $\psi$ is a small pair correlation due to proximity with the superconductors. Neglecting the higher order nonlinear term, at 
$\bm{B} = 0$, for a 1D with the exchange field in the $z$ direction, $\psi(x) = |\Delta|e^{i\alpha hx/\gamma}[e^{i\phi_R}\sinh \kappa(x + L/2) - e^{i\phi_L}\sinh \kappa(x - L/2)]/\sinh\kappa L,$
with $\kappa^2 = a/\gamma - \alpha^2h^2/\gamma^2$ and $\phi_{R/L} = \mp\left(\phi - \alpha h L/\gamma\right)/2$. 
Transparent boundary conditions have been assumed, i.e., $\psi(\pm L/2) = |\Delta|e^{i\phi_{R/L}}$, where $|\Delta|$ and $\phi$ is the absolute value of the superconducting gap, assumed equal in the 
two S, 
and their phase, respectively. Inserting $\psi(x)$ into \cref{eq:GLEL2} one finds the current-phase relation
\begin{align}
j = j_c\sin\left(\phi - \phi_0\right), 
\end{align}
where $j_c = \kappa|\Delta|^2/\sinh\kappa L$ and $\phi_0 = \alpha h L/\gamma$. The 
SOC has introduced a phase shift into the conventional Josephson current. This will be discussed in-depth later in this review.

Equation~(\ref{eq:GLEL2}) also reveals spontaneous edge currents in S/F structures, as noted in 
\onlinecite{Mironov2017}. If the interfacial SOC is substantial, $\bm{j}$ may be non-zero even if the orbital effect is negligible ($\bm{A} = 0$), and $\psi$ is uniform. In this case, $\bm{j}$ is directed along $\bm{n}\times\bm{h}$, parallel to the interface. This has interesting applications e.g., in a superconducting loop in proximity to  a ferromagnetic insulator, circulating spontaneous supercurrents are predicted~\cite{Robinson2019}, which may find use in single flux quantum (SFQ) logic---similarly to proposals involving $\pi$ junctions~\cite{Feofanov2010}. Superconducting vortices, generated %
due to these spontaneous currents, without an applied external magnetic field, have also been predicted at 
S/F interfaces~\cite{OldeOlthof2019}.

\subsubsection{Bogoliubov-de Gennes method}\label{bdg}
In the following, we will consider a superconducting system within the mean-field approximation. This can be described by a Hamiltonian of the form
\begin{align}
H =& \sum_{ss'}\int d\bm{r}\;\psi^\dagger_s(\bm{r})h_{ss'}(\bm{r})\psi_{s'}(\bm{r}) \nonumber\\
 +& \frac{1}{2}\int d\bm{r}\; \left[\Delta(\bm{r})\psi^\dagger_\uparrow(\bm{r})\psi^\dagger_\downarrow(\bm{r}) +\Delta^*(\bm{r})\psi_\downarrow(\bm{r})\psi_\uparrow(\bm{r}) \right], 
\label{eq:HBdG0}
\end{align}
where $\psi_s(\bm{r})$ is the field operator for an electron with spin $s$, $h$ is the single particle Hamiltonian containing SOC \cite{simensen_prb_18}, and in a homogeneous system $\Delta$ is the $s$-wave superconducting gap. 
The presence of $\Delta$ introduces the added complication that the electron and hole bands hybridize,  
described by using the Nambu basis $\Psi(\bm{r}) = \begin{pmatrix} \psi_\uparrow(\bm{r}) & \psi_\downarrow(\bm{r}) & \psi^\dagger_\uparrow(\bm{r}) & \psi^\dagger_\downarrow(\bm{r})\end{pmatrix}^T,$
in which case \cref{eq:HBdG0} may be written as
\begin{align}
H = H_0 + \frac{1}{2}\int d\bm{r}\; \Psi^\dagger(\bm{r})\hat{H}\Psi(\bm{r}),
\label{eq:BdGc}
\end{align}
where $H_0$ describes a trivial energy shift, and
\begin{align}
\hat{H} = \begin{pmatrix} h & \Delta i\sigma_y \\ -\Delta^* i\sigma_y & -h^*\end{pmatrix}.
\label{eq:Hbdg}
\end{align}
The diagonalization of \cref{eq:Hbdg} is referred to as Bogoliubov-de Gennes method~\cite{DeGennesBook}, where the quasiparticles (the eigenvalues) become a mixture of particles and holes.

Generally, two approaches are used when studying superconducting hybrid structures using the Bogoliubov-de Gennes method. A continuum formulation may be used, in which case the scattering at interfaces between materials is taken into account via generalizing the Griffin-Demers or Blonder-Tinkham-Klapwijk (BTK) formalism~\cite{Griffin1971:PRB,Blonder1982}. This entails matching the wave functions obtained from \cref{eq:BdGc} in adjacent materials at every interface and includes both normal reflection and tunneling processes, as well as Andreev reflection of opposite or equal 
spins~\cite{Zutic1999:PRBa}. While BTK formalism
assumes a step-function profile for the pair potential, the continuum formulation can also be solved self consistently~\cite{Valls:2022,Halterman2015:PRB,Setiawan2019:PRB,Wu2018:PRB}.

The other way of applying the Bogoliubov-de Gennes method is in a tight-binding approach on a lattice \cite{Bogoliubov-book}. It is appropriate when one wishes to study the equilibrium properties of superconducting systems---for instance the superconducting pair correlation in hybrid structures due to the proximity effect. The Hamiltonian then becomes a discrete $4N\times4N$ matrix, where $N$ is the number of lattice sites. An advantage of the Bogoliubov-de Gennes method is its conceptual simplicity, the drawback being the limitation in system size that is manageable computationally. It is a microscopic theory valid at arbitrary temperature.

\subsubsection{Quasiclassical theory}\label{sec:qc}
The Green function method is a powerful tool to describe condensed matter systems. Here, we briefly review the Keldysh technique, which is applicable to equilibrium and nonequilibrium systems. For an in-depth discussion, see~\cite{Rammer1986}. In many systems there is a dominating energy scale, such as the Fermi energy  $E_{\text{F}}$ in metals, so that all other contributions to the Hamiltonian may be considered small in comparison. In that case, and for metals in particular, the relevant contribution to several physical quantities of interest comes from near the Fermi level---referred to as the low-energy region. The Gor'kov equations, on the other hand, contain information about the entire spectrum. 
The quasiclassical approximation simplifies these equations by retaining only their low-energy component, i.e., terms which are at most linear in $\Xi/E_{\text{F}}$, where $\Xi$ is any of the {energy/self-energy scales} involved, $\Xi\in\left\{ |\Delta|, |\bm{h}|, \alpha,\ldots\right\}$
~\cite{Serene1983,Rammer1986,Millis1988:PRB,Belzig1999,Chandrasekhar2004}. 

The quasiclassical theory for SOC was established in~\onlinecite{Vorontsov2008:PRL, gorini_prb_10, raimondi_annal_12, Eschrig:2012}. Linear-in-momentum models of SOC such as the Rashba and Dresselhaus models may be introduced by an effective SU(2) gauge field, for which the derivative operator may be replaced by its gauge covariant equivalent. Within the quasiclassical approximation, it is given as 
$\tilde{\nabla} \bullet = \nabla \bullet - (ie/\hbar)\left[\bm{A}\;,\; \bullet\right],$ when acting on a $2\times 2$ Green function matrix $\bullet$ in spin space, where $A_k = A_{0,k}\sigma_0 + \alpha_{klm} \sigma_l k_m$, with $A_{0,k}$ a scalar gauge field stemming from a potential external magnetic field, and $\alpha_{klm}$ a generic tensor describing the SOC. For ballistic systems, the SOC has the form of a 
momentum-dependent exchange field, which is intuitively reasonable, and has interesting consequences in \eg ballistic Josephson weak links~\cite{Konschelle2016}.

The quasiclassical approximation of the Gor'kov equation of motion for the Green function of a ballistic system is the Eilenberger equation~\cite{Eilenberger}, and takes the form
\begin{align}
\hbar \bm{v}_{\text{F}}\cdot\tilde{\nabla}\check{g} + i\left[\varepsilon\check{\tau}_z + \check{\Sigma}\;,\;\check{g}\right] = 0,
\label{eq:eilenberger}
\end{align}
to lowest order where quasiparticle energy $\varepsilon$ is the Fourier conjugate to $t-t'$, assuming a stationary system, $\bm{v}_{\text{F}}$ is the Fermi velocity, $\check{\Sigma}$ contains the self energies and molecular fields acting on the itinerant electrons (such as a Zeeman-field) under study. Furthermore, $\check{\ldots}$ denotes an 8$\times$8 matrix in Nambu-Keldysh space, and $\check{g} = \frac{i}{\pi}\int d\xi\;\check{G}$ is the quasiclassical Green function, with $\xi = \hbar^2\bm{k}^2/2m - \mu$. Uniqueness of \cref{eq:eilenberger} is assured by the accompanying constraint $\left(\check{g}\right)^2 = \hat{I}$, with $\hat{I}$ the $4\times4$ identity matrix~\cite{Shelankov}.

A high concentration of impurities may also be treated within the quasiclassical approximation, using conventional impurity averaging techniques~\cite{Abrikosov1975}. This has the effect that quasiparticle motion takes the form of a random walk due to frequent impurity scatterings, so that 
momentum-dependent effects are strongly suppressed. The equation of motion then becomes
\begin{align}
D\tilde{\nabla}\cdot\check{g}\tilde{\nabla}\check{g} + i\left[\varepsilon\check{\tau}_z + \check{\Sigma}_\text{diff} + \bm{h}\cdot\check{\bm{\sigma}}\;,\;\check{g}\right] = 0,
\label{eq:usadel}
\end{align} 
where $D$ is the diffusion constant. The self-energy term $\check{\Sigma}_\text{diff}$ differs from $\check{\Sigma}$ in \cref{eq:eilenberger} in that the non-magnetic impurity potential that appears in the latter in a diffusive system has been averaged out. We have also separated a potential exchange (or Zeeman) field $\bm{h}$ from $\check{\Sigma}_\text{diff}$, as such a field is typically necessary to obtain spin-dependent effects due to the lowest-order contribution from SOC in the diffusive limit.  
Equation~(\ref{eq:usadel}) takes the form of a diffusion equation, and a net particle current, representing a drift in the random walk, may be identified by Fick's first law, $\check{\bm{j}} = -D\check{g}\tilde{\nabla}\check{g}$. 
This drift has a profound effect on diffusive systems, 
impurity scattering becomes anisotropic, and gives rise to momentum-dependent effects which cancel out in an isotropic system. 

Information about the superconducting correlations is found from the anomalous Green function $f$, which are off-diagonal $2\times 2$ blocks extracted from $\check{g}$, see, e.g.,~\onlinecite{Belzig1999} for a more detailed discussion. The effect of SOC on diffusive superconducting hybrid structures~\cite{bergeret_prl_13,bergeret_prb_14} is most evident in the limit where the proximity effect is weak, so that $f$ is small. By parameterizing into singlet $f_s$ and triplet $\bm{f}$ components using the $d$ vector notation~\cite{Leggett1975}, $f = \left(f_s + \bm{f}\cdot\bm{\sigma}\right)i\sigma_y$, one finds that for Rashba SOC, $\bm{A} = \alpha \bm{n}\cdot\left(\nabla\times\bm{\sigma}\right)$, where $\bm{n}$ is a unit vector pointing in the direction of symmetry breaking, the Usadel equation becomes $D\nabla^2f_s + 2i\varepsilon f_s + 2i\bm{h}\cdot\bm{f} = 0$, 
$D\nabla^2\bm{f} + 4iD\alpha\left[\bm{n}\times\left(\nabla\times\bm{f}\right) - \nabla\times\left(\bm{n}\times\bm{f}\right)\right]
+ \left(2i\varepsilon -4D\alpha^2\right)\bm{f} + 2i\bm{h}f_s = 0.$
A similar set of equations may be derived for the linear Dresselhaus model. Inspection of the above reveals that the singlet superconducting correlation is independent of the SOC to this level of accuracy, which is reasonable as it is independent of spin. The triplets, on the other hand, are influenced. The term proportional to $\alpha^2$ in the above equation is present also in homogeneous systems, and is a consequence of the frequent impurity scatterings in the diffusive limit. As momentum direction is averaged out by these scattering processes, so too is the spin direction---because of the SOC. This is an effect akin to the Dyakonov-Perel spin relaxation~\cite{Dyakonov1971:PL}, and leads to an increased decay of the triplet superconducting correlations. The terms proportional to $\alpha$ represent a precession of the pair correlation spins, resulting in a mixing of the triplet components. This observation, along with the experimental accessibility of the diffusive limit, has led to these systems receiving significant attention~\cite{Alidoust2015,Alidoust2015a,Jacobsen2015,Jacobsen2016, Amundsen2017}. The formalism has also been extended to incorporate magnetoelectric effects and particle-hole asymmetry ~\cite{Konschelle2015,Tokatly2017,Bobkova2017a, silaev_prb_17, virtanen_prb_22}, as well as extrinsic SOC, i.e., as induced by impurities~\cite{bergeret_prb_16,huang_prb_18,Virtanen2021}. This extension allows for the description of additional physical effects such as anomalous Josephson currents and spin Hall effects in the superconducting state. Recent works have reconciled the nonlinear %
quasiclassical equations including SOC with a normalized Green function \cite{Virtanen2021, virtanen_prb_22}, and also expanded the theory to include boundary conditions describing interfacial SOC \cite{Amundsen2019}, giving rise to a supercurrent spin Hall effect \cite{linder_prb_22}.

\section{Experimental techniques}

Here, we provide some intuitive guidelines on designing experiments to investigate S/F proximity effects in the context of SOC-driven triplet pair formation. Although the techniques are similar to standard S/F proximity effects, we highlight, where appropriate, the expected differences in the outcomes when considering the triplet pair formation due to SOC.

\subsection{Transition temperature measurements}

Measuring $T_c$ of a thin film S proximity-coupled to F layers is a common method of exploring S/F structures. The S/F proximity depends on the depairing effect of the magnetic exchange field on the Cooper pairs and/or the generation of triplet pairs affecting the singlet pairing amplitude in S. 

The majority of $T_c$ measurements are carried out on unpatterned thin films using a four-point current-bias technique. A low bias current is used to avoid current-induced nonequilibrium shifts in $T_c$. To control the magnetic state of the F layer(s), magnetic fields are applied which are negligible relative to the upper critical field of S layer. There is an orbital depairing effect from the applied magnetic field which can suppress $T_c$ and must be accounted for during the analyses of the proximity effect. This is usually not a problem for IP magnetic fields as the coercive fields of transition metal Fs are small relative to the IP upper critical field of the S layer. However, the situation is more complex for OOP magnetic fields where the coercive fields of F layers tend to be large and are often comparable to or greater than the magnetic field required to nucleate superconducting vortices. Further complication arises from the dipolar fields injected into S from magnetic domain walls. These factors should be considered during the analyses of the results and often requires control samples including isolated S films and S/F structures with insulating barriers to break the proximity effect between S and F layers (e.g.,~\cite{Banerjee2018, Singh:2015}.

Careful consideration is required when selecting the F material. For S/F/F' or F/S/F' spin valves, the key challenge is to obtain stable parallel, antiparallel or noncollinear magnetic states over a range of reasonable magnetic fields. Here, F and F' refer to two distinct F layers, with sufficiently different coercive fields. 
This requires specific anisotropies e.g., Co/Ni multilayers~\cite{satchell_prb_19} or applying a large OOP field to reorient one F layer (e.g. Ni)~\cite{Singh:2015}. 

\subsection{Josephson junctions}

JJs with F barriers have been key to demonstrating triplet creation with one of the first experiments detecting supercurrents through the highly spin-polarized half-metallic CrO$_2$~\cite{keizer_nature_06}. The absence of minority spin states in CrO$_2$ means that any supercurrent flowing through it must be mediated by spin-charge triplet current; however, magnetic control of singlets to triplet pair formation was found to be challenging and highly irreproducible. Since then advances in triplet supercurrent transport in JJs has been made as  discussed in Section IV E. 

Owing to the small electrical resistance of metallic S/F heterostructures, nanopatterning is required for straightforward measurements of device voltage e.g., in JJs. For JJs with magnetic barriers, a further advantage of nanopatterning is that the F layers can be magnetically single domain meaning that even at high applied magnetic fields, the barrier flux can be small relative to a flux quanta. This allows manipulation of the magnetic state of the barrier without significantly lowering the JJ 
critical current. 
The presence of a barrier magnetic moment adds to the magnetic flux from the external magnetic field and can distort the magnetic-field-dependence of the Josephson critical current. A further complication in nanopatterned JJs is that dipolar fields from the F layers can distort the single domain state, introducing additional complex nanomagnetic states. These issues need careful consideration when designing JJs with magnetic barriers~\cite{Blamire_2013}. 

Optical or electron beam lithography techniques are routinely used for fabricating S/F devices, including JJs. These techniques are described by~\onlinecite{Blamire_2011_nanopillar}. 

\subsection{Magnetization dynamics}

An injected normal state current from a thin film F into an S in S/F structures can introduce a nonequilibrium quasiparticle spin accumulation with relaxation lengths much larger than in normal metals~\cite{Parkin_2010_NM}, differing also from the charge relaxation length~\cite{Charis_2013_N.Phys,Hubler_12}. This has been explained by the Zeeman splitting of the quasiparticle bands, combined with an energy imbalance~\cite{Silaev2015:PRL,Bergeret2018:RMP,Bobkova2015:JETP}.  Alternatively, following the original experiment on spin dynamics of ferromagnets in proximity to superconductors \cite{Bell_2007_PRL}, it may be possible to create a superconducting spin current in the S layer mediated via triplet pairs~\cite{Jeon:2019}. This experiment involved spin pumping from a layer of NiFe (permalloy, Py) in a  Pt/Nb/Py/Nb/Pt structure. Here, the effective Gilbert damping, $\alpha$, which is proportional to the spin-current density, of the precessing {\bf M} of Py increased below $T_c$, indicating an enhancement of the spin current above the normal state which is different from the spin pumping via the Andreev bound states~\cite{Yao_2021_spinpumping}.

Spin pumping in S/F structures can be performed using broadband ferromagnetic resonance (FMR). For FMR experiments on S/F/S structures, the S layer thickness should be 
below the magnetic penetration depth ($\sim$100 nm for Nb), otherwise the dc-resonance field shifts to lower values ~\cite{PhysRevApplied.11.014061,Li_2018,PhysRevApplied.14.024086}. 
\onlinecite{PhysRevApplied.11.014061}
attribute this to Meissner screening fields generated in the S layer. An alternative explanation by \onlinecite{PhysRevApplied.18.L061004} suggests that a gap is spontaneously induced in the magnon spectrum through the Anderson-Higgs mechanism, causing a shift in the dc-resonance field of a S/F/S structure.

\subsection{Spectroscopic techniques}

The nature of a triplet-induced in a F, S or normal metal (N) layer is theoretically different from the usually dominant singlet state. Although odd-frequency triplet and singlet states share the common feature of an $s$-wave order parameter, unlike the singlet state, the orbital and spin components of the triplet state are even with respect to the electron exchange and this implies their correlation function must be odd in frequency or, equivalently, odd under exchange of time coordinates \cite{bergeret_prl_01, linder_rmp_19}. Consequently, an odd-frequency triplet state can enhance the quasiparticle density of states (DOS) and conductance of an S/F bilayer~\cite{Kontos_2002_PRL,Petrashov_1999_PRL}. This can be shown by considering the Green function for an odd-frequency superconductor at small energies, where it looks like a normal metal with an effective mass renormalization (increase), enhancing the DOS \cite{linder_rmp_19}. For a singlet state, there is no enhancement in the DOS and so the observation of a conductance peak around zero voltage is an indication of odd-frequency superconductivity. Such quantized peak was predicted as a signature of Majorana zero modes~\cite{Sengupta2001:PRB} (see Sec.~IVA), 
but alternative explanations are possible~\cite{Yu2021:NP,DasSarma2021:PRB}.

The standard technique to probe the induced superconducting DOS involves tunnel junctions or scanning tunneling microscopy measurements of the current-voltage, $I(V)$, characteristics. Below $T_c$, the $I(V)$ characteristics are nonlinear at voltages near the gap edge. 
For S/insulator/N junctions that is
around $V = \pm\Delta/e$, where $e$ is the electron charge and $\Delta$ is the energy gap. Such measurements can provide high-resolution spectroscopic information because $dI/dV$ is proportional to the quasiparticle excitations, i.e., to the superconducting DOS. Tunneling studies have been extensively used to probe the singlet state~\cite{SanGiorgio_2008_PRL,Boden_2011_PRB} and the triplet state~\cite{Kalcheim_2012_PRB,Kalcheim_2014_PRB,Kalcheim_2015_PRB} in different S/F structures.

\subsection{Low-energy muon spin rotation technique}

LE-$\mu$SR (or low-energy
muon spin rotation) has been used to probe the depth-dependence of superconductivity and magnetism in S/F structures. LE-$\mu$SR offers a high sensitivity to magnetic fluctuations and spontaneous fields $< 0.1$ G with a depth-resolved sensitivity of a few nanometers~\cite{Bernardo_2015_PRX,Fittipaldi2021, Flokstra_PRB_2014}.

Muons are spin-half elementary particles with a charge matching an electron but over two-hundred times heavier. Implanted muons can provide detailed information about their local magnetic environment within a material~\cite{Hillier_NatRev_2022} and so muon spin rotation technique is a sensitive tool for probing sub-surface superconductivity in the Meissner state. Spin-polarized positive muons, $\mu^{+}$, are generated from $\pi^{+}$ decay and are moderated by passing them through a cryosolid, typically Ar, to obtain $\mu^{+}$ in the low-energy range ($\sim$15 eV). These $\mu^{+}$ are accelerated by an adjustable sample bias that tunes their energies from 0.5-30 keV, for their precise implantation depth within a material. The spins of the implanted $\mu^{+}$ precess about a local magnetic field and a decaying $\mu^{+}$ emits a positron in the direction of the $\mu^{+}$ spin. This decay of the positron intensity is measured as a difference in the number of counts by two detectors placed near the sample. 

To study the Meissner state of S or S/F structure, a magnetic field, $B_\mathrm{ext}$ is applied parallel to the sample plane and perpendicular to the initial spin polarization of the $\mu^{+}$ beam. This induces a precession of the $\mu^{+}$ spins at an average frequency of $\bar{\omega}_{s}=\gamma_{\mu}\bar{B}_{loc}$ where $\bar{B}_{loc}$ is the average local field experienced by the implanted muons and $\gamma_{\mu}=2\pi\times135.5$ $\mathrm{MHz T^{-1}}$ is the gyromagnetic ratio of the $\mu^{+}$. If the stopping distribution is $p(z,E)$ at a depth $z$ and energy $E$ of the implanted muons, then the precession frequency is $\bar{\omega}_{s}=\gamma_{\mu}\int{B}_{loc}(z)p(z,E)\,dz$. The asymmetry spectrum $A_s(t,E)$, which measures the normalized difference in the counts of the left and right detectors is proportional to $e^{-\bar{\lambda}t}[\cos{\gamma_{\mu}\bar{B}_{loc}t+\phi_{0}t}]$ for a given implantation energy $E$. Here, $\bar{\lambda}$ is the mean muon depolarization rate and $\phi_{0}E$ the starting phase of the muon precession. A series of mean-field values $\bar{B}_{loc}$ is determined from the asymmetry fits as a function of the muon implantation energy $E$ which provides the final $B_{loc}(z)$ profile inside the sample.

In the context of S/F proximity effects and
triplet pairs, this technique was used to detect a paramagnetic Meissner effect in Au/Ho/Nb~\cite{Bernardo_2015_PRX}, enhanced flux expulsion in Cu/Nb/Co~\cite{Flokstra_PRL_2018}, remotely induced magnetism in a normal metal in a superconducting spin valve structure~\cite{Flokstra_NPhys_2015} and Au/C$_{60}$/Cu/C$_{60}$/Nb~\cite{Rogers2021_ComPhys} structures, observed as a local enhancement of the 
magnetic field in Au above the externally applied field, in contrast to the well-known Meissner effect in the superconducting state. In the presence of SOC, the Meissner response has been predicted to be tunable from paramagnetic to diamagnetic \cite{espedal_prl_16}.

\section{Recent developments}
In this section we summarise the recent theoretical and experimental developments with a specific motivation to show how triplet pairing induced by SOC provides a common link to a wide range of phenomena - from Majorana zero modes to superconducting spintronics.

\subsection{Majorana zero modes}
Majorana fermions are particles which have the peculiar property of being their own anti-particle. 
They are real solutions of the Dirac equation and represent a potential new, as of yet, undetected fundamental particle~\cite{Elliott2015:RMP}. In condensed matter systems, predicted Majorana fermions 
are chargeless quasiparticle excitations~\cite{Aguado2017:RNC}. This property makes superconductors ideal candidates to host such states, as the quasiparticles of superconducting systems---the bogoliubons---can consist of an equal mixture of the electron-like and hole-like excitations of the normal state system, as discussed in section \ref{bdg}.
However, these superconductors cannot be conventional 
with a spin-singlet configuration. 
Instead, topological superconductors are sought
with equal-spin (also referred to as spinless) pairing which, as topological insulators, feature a band inversion and a nontrivial topology~\cite{Culcer2020:2DM,Shen:2012}. 
Defects (such as vortices) and quasiparticles in topological superconductors or boundaries between topological and trivial regions, can bind localized Majorana zero modes (MZM)~\cite{Aasen2016:PRX}. These zero-energy (pinned at the Fermi level) topologically-protected degenerate states, in which quantum information can be nonlocally stored, are separated by the topological gap from the excited states. 

\begin{figure*}
\resizebox{18cm}{!}{\includegraphics{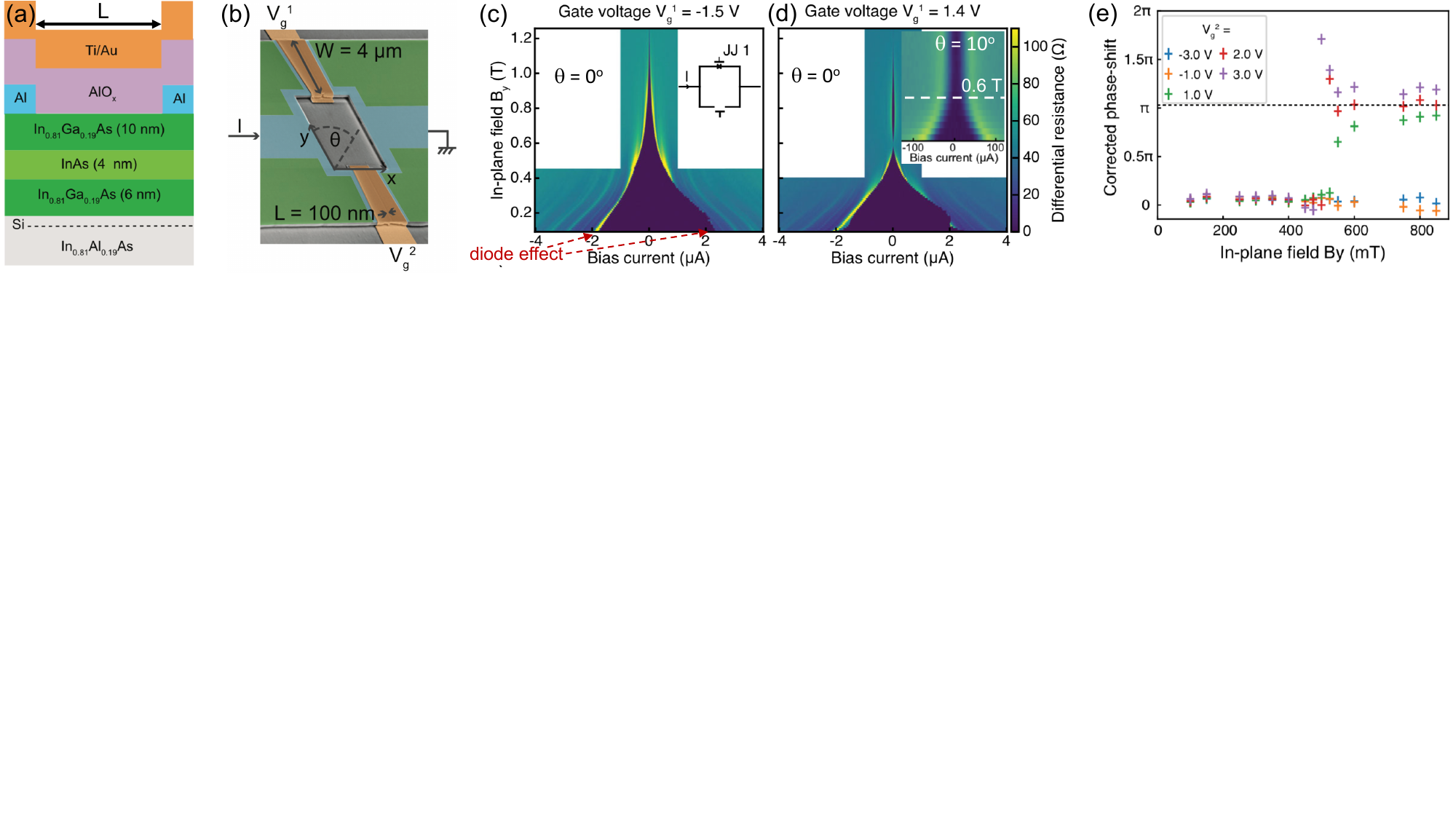}} 
\vspace{-7.1cm}
\caption{Experimental evidence for topological superconductivity. (a) Schematic diagrams of a planar JJ. (b) SEM micrograph of a SQUID formed of two JJs with width W$ = 4~\mu$m and separation between the S contacts (Al) L$ = 100~$nm. Each JJ is independently gated with the voltage tuning both the carrier density and the SOC.  
The x-direction is colinear to the current flow in the JJs. Differential resistance of JJ1 as function of an applied in-plane field at gate voltages: (c) V$^1_g = -1.5~$V, (d) V$^1_g = 1.4~$V. In both cases, JJ2 is depleted (V$^2_g = -7~$V) and does not participate in the transport. Marked asymmetry signifies the superconducting diode effect.
In (d), as expected for the transition to topological superconductivity, a minimum of the critical current is observed around 0.6~T for JJ1 for the in-plane field, B$_y$, at $\theta = 0^o$ [see Fig.~\ref{fig:MZM_JJ}(b)]. Inset: For $\theta = 10^o$ such minimum is lost. (e) Phase signature of topological transition from SQUID interferometry. Phase shift between the SQUID oscillation at V$^2_g = -4~$V and the oscillation at a different value as a function of B$_y$.  The linear B$_y$-contribution (due to anomalous Josephson effect) has been subtracted to highlight the phase jump of  $\sim \pi$ at three higher V$^2_g$ values. From \onlinecite{Dartiailh2021:PRL}.
}
\label{fig:MZM_JJ}
\end{figure*}

A huge interest in MZM comes from their exotic non-Abelian statistics (unlike the Majorana fermions in particle physics), which is both fundamentally exciting and offers prospect for fault-tolerant topological quantum computing~\cite{Ivanov2001,Kitaev2003,Nayak2008b,Alicea2011,Sarma2015}. An interchange of the position of two MZM, known as braiding, yields a non-Abelian phase and transforms quantum states within a degenerate ground-state manifold. The result is a quantum gate, topologically-protected from local perturbations~\cite{Lahtinen2017:SPP} 
that limit conventional quantum computers.
A complementary signature of the non-Abelian statistics comes from bringing together, or fusing, two MZM which removes their degeneracy and yields either an ordinary fermion or a vacuum state (Cooper pair condensate)~\cite{Beenakker2020:SPP}. Experimental reports of MZM detection remain debated and do not include their non-Abelian statistics. 

In 1D, MZM are found at the ends of the Kitaev chain~\cite{Kitaev2001}, with triplet $p$-wave superconductivity. In 2D, MZM appear in $p_x \pm ip_y$ superconductors as localized states bound to vortices, as well as distributed chiral edge states at interfaces~\cite{Read2000}. 
Higher-order topological superconductors, analogous to their insulating counterparts~\cite{Benalcazar2017,Schindler2018,Schindler2018a}, can host MZM on surfaces of codimension larger than one, e.g., in the 0D 
corners of 2D systems, or 1D hinges of 3D systems.

There are materials believed to exhibit intrinsic topological superconductivity, such as Sr$_2$RuO$_4$~\cite{Mackenzie2003,Kallin2012},
UTe$_2$~\cite{Shishidou2021:PRB}, and  Cu$_x$Bi$_2$Se$_3$~\cite{Sasaki2011,Kriener2011}. 
However, challenges in relying on them are seen from the extensively studied Sr$_2$RuO$_4$. While doubts about the claimed $p_x \pm ip_y$~\cite{Mackenzie2003,Kallin2012,Nelson2004:S} were raised before~\cite{Zutic2005:PRL}, an experimental evidence against the $p$-wave~\cite{Petsch2020:PRL} now even involves the discoverer of Sr$_2$RuO$_4$~\cite{Maeno1994:N}.

An alternative to the elusive intrinsic $p$-wave superconductors, is using the proximity-induced topological superconductivity.  
Chiral triplet superconducting correlations can be induced on the semimetallic surface states of a topological insulator in proximity to a conventional $s$-wave superconductor~\cite{Fu2008:PRL,Rosenbach2021:SA}. The proximity-induced topological superconductivity is also predicted when the topological insulator is replaced by a semiconductor nanowire with strong Rashba SOC and a Zeeman field~\cite{Sau2010,Alicea2010,Lutchyn2010,Oreg2010:PRL,Brouwer2011:PRB,Mourik2012,Das2012,Deng2012,Rokhinson2012:NP}. For orthogonal Rashba SOC and Zeeman
field (Figs.~3 and 4), the spin degeneracy at $k=0$ is removed. The resulting spin canting and depletion of one spin 
support %
$p$-wave triplet topological superconductivity
with the effective gap $\Delta_p(k) \propto \alpha k \Delta/2\sqrt{h^2+\alpha^2k^2}$, when $\mu$
is %
within the Zeeman gap, $h$~\cite{Aguado2017:RNC,Lutchyn2010,Oreg2010:PRL}.

With strong SOC, noncentrosymmetric superconductors \cite{tanaka_prl_10, yada_prb_11} lead to both $p$-wave and $s$-wave superconducting correlations. 
These materials support topologically-nontrivial state with
edge modes~\cite{Tanaka2009,Smidman2017} and generate tunable higher-order topological states by rotating an IP magnetic field, $B_{||}$, to switch 
between corner and edge modes~\cite{Zhu2018,Pahomi2020,Ikegaya2021}. MZM are reported using a huge Rashba SOC ($\sim 110\,$meV) in Au(111) surface 
states~\cite{Wei2019:PRL,Manna2020:PNAS}.

Instead of native SOC, MZM can be hosted in systems with 
synthetic SOC realized through magnetic textures and the resulting fringing fields~\cite{Klinovaja2012:PRL,Kjaergaard2012:PRB,Nadj-Perge2014:S,Fatin2016:PRL,Kim2018:SA,Kim2015:PRB,Desjardins2019:NM,Gungordu2022:JAP,Wu2017:PRB,Marra2017:PRB, steffensen_prr_22, huang_prr_23}. Magnetic textures, ${\bf B}({\bf r})$, remove
the need for an applied magnetic field and their tunability reconfigures regions that
support topological superconductivity to create, control, and braid MZM~\cite{Fatin2016:PRL,Yang2016:PRB,Boutin2018:PRB,Zhou2019:PRB,Mohanta2019:PRA,Gungordu2018:PRB,Matos-Abiague2017:SSC}.  The emergent SOC is understood by noting that Zeeman interaction, $g_{\mathrm{eff}} \mu_B {\bf B}({\bf r})/2\cdot {\bf{\sigma}}$, 
where $g_{\mathrm{eff}}$ is the effective $g$-factor, is diagonalized by performing local spin rotations aligning the spin-quantization axis to the local ${\bf B}({\bf r})$, known for over
45 years~\cite{Matos-Abiague2017:SSC}. In a rotated frame,
the Zeeman energy is simplified, $|g_{\mathrm{eff}} \mu_B {\bf B}({\bf r})/2|\sigma_z$, while the kinetic energy acquires an extra term due to
the non-Abelian field that yields the synthetic SOC. 

Since the role of magnetic textures would be more pronounced in 
proximitized materials with large $|g_{\mathrm{eff}}|$, such as narrow-band semiconductors or
magnetically-doped semiconductors~\cite{Fatin2016:PRL,Zhou2019:PRB,Mohanta2019:PRA}, it was surprising that %
MZM were reported in a carbon nanotube~\cite{Desjardins2019:NM,Yazdani2019:NM}, where the weak inherent SOC renders $|g_{\mathrm{eff}}|$ small.
Experimentally, 
a multilayer Co/Pt magnetic textures generated strong fringing field $\approx
0.4$~T in the nearby carbon nanotube which resulted both in the Zeeman interaction and  
the characteristic SOC energy $\approx 1.1$~meV~\cite{Desjardins2019:NM}, exceeding the SOC values for InAs or InSb, common MZM 
candidates. Tuning the magnetic textures, which needs to be accurately studied through micromagnetic 
simulations~\cite{Zhou2019:PRB,Mohanta2019:PRA,Desjardins2019:NM}, 
revealed through the oscillations of the superconductivity-induced subgap states in carbon nanotube-based JJs with $s$-wave 
Pd/Nb electrodes.

Experiments in 2DEGs 
have revealed a strong proximity effect when coupled to a superconductor, even in the presence of strong SOC~\cite{Wan2015,Shabani2016,Kjaergaard2016}. 
Planar JJs with $B_{||}$, where the superconducting correlations in the 2DEG can be tuned into a topologically nontrivial phase by a phase difference between the superconducting banks, are predicted to produce MZM~\cite{Pientka2017,Hell2017,Hell2017a,Stern2019,zhou2020:PRL} and accompanied
by related experiments reporting topological superconducitivity~\cite{Fornieri2019:N,Ren2019:N,Dartiailh2021:PRL}. Using a
SQUID geometry as shown in Figs.~\ref{fig:MZM_JJ} (a) and (b), 
gate voltage can change both the carrier density and the strength of the Rashba SOC, which can determine
the presence or absence of the topological superconductivity. Two individually gated JJs in the SQUID thus control the current to flow through both or just one of them~\cite{Dartiailh2021:PRL}. Figures~\ref{fig:MZM_JJ}(c) and (d) show that with gate control the JJ current can become nonmonotonic with $B_{||}$, as expected from the closing of the $s$-wave
and the reopening of the $p$-wave superconducting gap, predicted in proximitized 
nanowires~\cite{Sau2010,Alicea2010,Lutchyn2010,Oreg2010:PRL}. The observed JJ current anisotropy,
where its nonmonotonic character is lost for $B_{||}$ which is sufficiently
misaligned with the N/S interface [Fig.~\ref{fig:MZM_JJ}(d), inset], further 
supports the expected proximity-induced $p$-wave superconductivity. 

An independent signature of the topological superconductivity is obtained from the SQUID measurements which in Fig.~\ref{fig:MZM_JJ}(e) reveal an approximate $\pi$ jump in the superconducting phase difference, expected across the transition to topological superconductivity~\cite{Pientka2017,Hell2017}. These various signatures of the topological superconductivity are obtained on the same sample and indicate topological transition at $\sim 0.6~$T.
In contrast, $B_{||}$ required to reach $0-\pi$ transition expected from
the FFLO-like mechanism~\cite{Yokoyama2014:PRB} in the studied 
samples is 
$B_{0-\pi}=(\pi/2) \hbar v_\mathrm{F}/(g \mu_\mathrm{B} L)\approx 14.4~$T~\cite{Dartiailh2021:PRL}. In another Al/InAs planar JJ, topological superconductivity was reported at an even lower  $B_{||} \sim 0.2\,$T~\cite{Banerjee2023:PRB}. With phase bias $\pi$, idealized JJs could host MZM even at $B_{||}=0$, but determining the 
optimal topological gap is complicated by crystalline and magnetic anisotropy and a finite geometry~\cite{Paudel2021:PRB,Pekerten2022:PRB}.
With multiple gates, planar JJs can be used to 
fuse MZM and probe the 
non-Abelian statistics~\cite{Zhou2022:NC}. 

Curved nanostructures, for next-generation spintronic devices~\cite{Nagasawa2013,Gentile2015,Ying2016,Chang2017,Francica2019,Das2019, salamone_prb_21, salamone_prb_22}, could host MZM. SOC,  
forcing motion along curved geometries can lead to nontrivial spin-dependent effects. Furthermore, bending, e.g., a nanowire, introduces a strain field which itself acts as a source of SOC. When superconducting order is introduced to such systems, curvature-dependent triplet superconducting correlations may appear~\cite{Ying2017}, a necessary ingredient for MZM. 
The manipulation of curvature can be used to exert control over, and even induce, nontrivial topology and MZM~\cite{Francica2020,Chou2021}.

\subsection{Superconducting critical temperature}
Some theory~\cite{Tagirov:1999, Buzdin:2001} and measurements of F/S/F trilayers with both weak (CuNi) and strong (NiFe) F layers showed that $T_c$ is higher for antiparallel F-layer moments versus parallel F-layer moments~\cite{Gu:2002,Moraru:2006}. This can be understood from the higher net pair-breaking exchange field for parallel F layer moments which strongly suppresses singlet  superconductivity. However, other experiments \cite{rusanov_prb_06} observed a higher $T_c$ in the P state, which can be partially understood from a suppression of inverse crossed Andreev reflection in the P state \cite{fominov_jetp_10, mironov_prb_14}. For noncollinear F-moment alignments in F/S/F and S/F/F systems~\cite{Leksin:2012,Wang:2014}, the proximity effect between the S and F layers is enhanced due to the generation of triplet Cooper pairs. The increased proximity effect results in a reduction of $T_c$ by up to 120 mK for $3d$ ferromagnets~\cite{Leksin:2012,Wang:2014} and 1 K for half-metallic ferromagnets such as CrO$_2$~\cite{Singh:2015}.

The principle of detecting triplets via a magnetic-state-dependent modulation of $T_c$ was experimentally used to detect control of short-ranged triplets~\cite{Banerjee2018} (see Section II C for a definition of short-ranged triplets). Using a Pt/Co/Pt trilayer proximity-coupled to an s-wave superconductor (Nb), a strong suppression of $T_c$ for magnetic fields applied IP and partial compensation of $T_c$ suppression for OOP fields was detected. This was in sharp contrast to a pure Nb or Nb/Co multilayers where relatively little $T_c$ suppression is seen for IP fields with negligible orbital depairing and a strong OOP $T_c$ suppression arising from orbital effects. The unconventional modulation is explained by the fact that in S/F structures without SOC, the short-ranged triplet energy does not depend on {\bf M} orientation thereby making the $T_c$ independent of
{\bf M} angle $\theta$ with the film plane. However, in presence of SOC arising from the interfacial symmetry breaking in Pt/Co/Pt trilayers, an increasing IP field increases the “leakage” of the Cooper pairs through the triplet channel. This leakage drains the superconductor of Cooper pairs and the superconducting gap is reduced. An OOP has an opposite effect and closes this parallel triplet channel, thereby reducing the $T_c$ suppression. The dependence of the magnitude of $T_c$ modulation is expected to depend both on the strength of the SOC and {\bf M} with the former most likely depending on the exact interface structure~\cite{Bregazzi2024}. Moreover, disorder will suppress odd-parity superconducting correlations, due to the randomization of electron momentum, generated by the lack of translational symmetry at the interface, which in turn will modify $T_c$.

\subsection{Modification of magnetic anisotropy} A consequence of a SOC-driven modulation of superconductivity is the potential for a reciprocal effect i.e., a reorientation of {\bf M} due to superconductivity~\cite{Johnsen:2019}. A reduction in $T_c$ of the superconductor for an IP {\bf M} translates to a reduction in the condensation energy due to a suppression of the superconducting gap. The free energy of the superconducting state will thus favour an OOP {\bf M}. {\bf M}-angle-dependence of free energy means that for a sufficiently low-anisotropy barier in the F layer, $T_c$ can trigger an IP to OOP {\bf M} reorientation. This is modeled as a S/HM/F structure. Using the tight-binding Bogoliubov-de Gennes method on a lattice (see Section 3), the system is described by the Hamiltonian 
\begin{equation}
    \begin{split}
        \label{Hamiltonian}
        &H = -t\sum_{\left<\boldsymbol{i},\boldsymbol{j}\right>,\sigma}c_{\boldsymbol{i},\sigma}^\dagger c_{\boldsymbol{j},\sigma}
        -\sum_{\boldsymbol{i},\sigma} \mu_{\boldsymbol{i}} c_{\boldsymbol{i},\sigma}^\dagger c_{\boldsymbol{i},\sigma}
        -\sum_{\boldsymbol{i}} U_{\boldsymbol{i}}n_{\boldsymbol{i},\uparrow}n_{\boldsymbol{i},\downarrow} \\
        &\hspace{8.5mm}-\frac{i}{2}\sum_{\left< \boldsymbol{i},\boldsymbol{j}\right> ,\alpha,\beta} \lambda_{\boldsymbol{i}} c_{\boldsymbol{i},\alpha}^{\dagger} \hat{n} \cdot (\boldsymbol{\sigma}\times\boldsymbol{d}_{\boldsymbol{i},\boldsymbol{j}})_{\alpha,\beta} c_{\boldsymbol{j},\beta}\\
        &\hspace{8.5mm}+\sum_{\boldsymbol{i},\alpha,\beta}c_{\boldsymbol{i},\alpha}^{\dagger}(\boldsymbol{h}_{\boldsymbol{i}}\cdot\boldsymbol{\sigma})_{\alpha,\beta} c_{\boldsymbol{i},\beta}
    \end{split}
\end{equation}
Here, $t$ is the hopping integral, $\mu_{\boldsymbol{i}}$ is the chemical potential at lattice site $\boldsymbol{i}$, $U<0$ is the attractive on-site interaction which gives rise to superconductivity, $\lambda_{\boldsymbol{i}}$ is the Rashba SOC magnitude at site $\boldsymbol{i}$, 
$\hat{n}$ is a unit vector normal to the interface, 
$\boldsymbol{d}_{\boldsymbol{i},\boldsymbol{j}}$ is the vector from site $\boldsymbol{i}$ to site $\boldsymbol{j}$, and $\boldsymbol{h_i}$ is the local magnetic exchange field.  $c_{\boldsymbol{i},\sigma}^\dagger$ and $c_{\boldsymbol{i},\sigma}$ are the second quantization electron creation and annihilation operators at site $\boldsymbol{i}$ with spin $\sigma$, and $n_{\boldsymbol{i}}\equiv c_{\boldsymbol{i},\sigma}^\dagger c_{\boldsymbol{i},\sigma}$. The superconducting term in the Hamiltonian is treated by a mean field approach, where  $c_{\boldsymbol{i},\uparrow} c_{\boldsymbol{i},\downarrow} = \left< c_{\boldsymbol{i},\uparrow} c_{\boldsymbol{i},\downarrow} \right> +\delta$ and $c_{\boldsymbol{i},\uparrow}^\dagger c_{\boldsymbol{i},\downarrow}^\dagger = \big< c_{\boldsymbol{i},\uparrow}^\dagger c_{\boldsymbol{i},\downarrow}^\dagger \big> +\delta^\dagger$ is inserted into Eq.~(\ref{Hamiltonian}) 
and neglect terms of second order in the fluctuations $\delta$ and $\delta^\dagger$. $\Delta_{\boldsymbol{i}} \equiv U_{\boldsymbol{i}}\left<c_{\boldsymbol{i},\uparrow}c_{\boldsymbol{i},\downarrow}\right>$ is the superconducting order parameter which is solved self-consistently. In presence of a strong shape anisotropy favouring an IP orientation, this model predicts a $\pi/4$ rotation in the plane of the film below the superconducting transition. At a rotation of $\pi/4$ with respect to the crystal axes, the triplet generation is minimised which maximises the superconducting condensation energy required for the {\bf M} rotation.

These predictions were recently demonstrated in magnetic tunnel junctions containing epitaxial V/MgO/Fe/MgO/Fe/Co~\cite{Gonzalez:2021} grown using molecular beam epitaxy. The Rashba SOC arises from the MgO/Fe interface which is also responsible for a well-defined perpendicular magnetic anisotropy
in addition to the required cubic symmetry of Fe(001). The top Fe/Co bilayer detects superconductivity-driven orientational changes of the Fe(001) layer through the tunneling magnetoresistance (TMR) effect. Below $T_c$ 
of the V layer, the Fe layer showed a pronounced reduction in the field required to orient the OOP {\bf M} or for larger junctions a spontaneous reorientation in the OOP direction at zero field. Interestingly, an electric field effect was also reported in the superconducting state where the OOP switching fields depends on the strength and direction of the applied field~\cite{Gonzalez:2021}. This field-dependent behaviour arises from an electric field-induced modification of the Rashba SOC. A similar effect, albeit IP rotation of the Fe(001) layer magnetic moment, was also observed in this system~\cite{Gonzalez:2020} as predicted by the theoretical model described above. While the free energy considerations are somewhat similar, this is fundamentally different from the superconducting exchange coupling observed in GdN/Nb/GdN trilayers~\cite{Zhu:2017} and originally predicted by \onlinecite{deGennes:1966}.

\subsection{Interfacial magnetoanisotropy}

TMR~\cite{Tsymbal:2019} is an important effect in spintronics where the tunneling probability, and thus the resistance, of F/insulator/F trilayers
depend on the orientation of the two ferromagnets. 
In a N/F bilayer 
with an  interfacial SOC 
the resistance can depend 
on the {\bf M} orientation from TAMR
~\cite{Gould2004:PRL,Moser2007:PRL}, recall section \ref{SOC}. TAMR devices have an  
advantage 
over TMR equivalents as they requires only a single F
thereby reducing the number of interfaces and potential alignment problems due to magnetostatic coupling between the two F layers. 

When the normal metal is replaced with a superconductor, Andreev reflection provides an additional source of magnetoanisotropy through a process known as magnetoanisotropic Andreev reflection (MAAR)~\cite{Hogl2015}. The transport properties of such a bilayer has been explored in~\cite{Vezin2020} using the BTK formalism outlined in section \ref{bdg}, within which the interfacial SOC may be included as a boundary potential. With the F/S interface located at 
$z = 0$, the Hamiltonian is $H = \hbar^2\bm{k}^2/2m - \mu + \bm{h}\cdot\bm{\sigma}\theta(z) + V_B(z),$
where $\bm{h}$ is the F exchange field, and $\theta(z)$ is the 
step function. The boundary potential is assumed to contain a spin-independent contribution, $V_0$, as well as Rashba SOC $V_B(z) = \left[V_0d + \alpha\left(k_x\sigma_y - k_y\sigma_x\right)\right]\delta(z),$
with $d$ the barrier thickness.
It is convenient to introduce the 
dimensionless quantities: spin polarization $P={|\bm{h}|/{(2\mu_F})}$, barrier %
$Z =V_0 d \sqrt{m}/(\hbar^2 \sqrt{k_F q_F})$,
and Rashba SOC $\lambda = 2\alpha \sqrt{m}/ \hbar ^2$ strengths, where $\mu_F$ is the chemical 
potential in the F region, $m$ the effective mass, while $k_F$ ($q_F$) is the magnitude of the (spin-averaged)
wave vector in the F (S) region. 
\begin{figure}[h]
\vspace{-1.5cm}
    \centering
    \includegraphics[width=13.7cm]{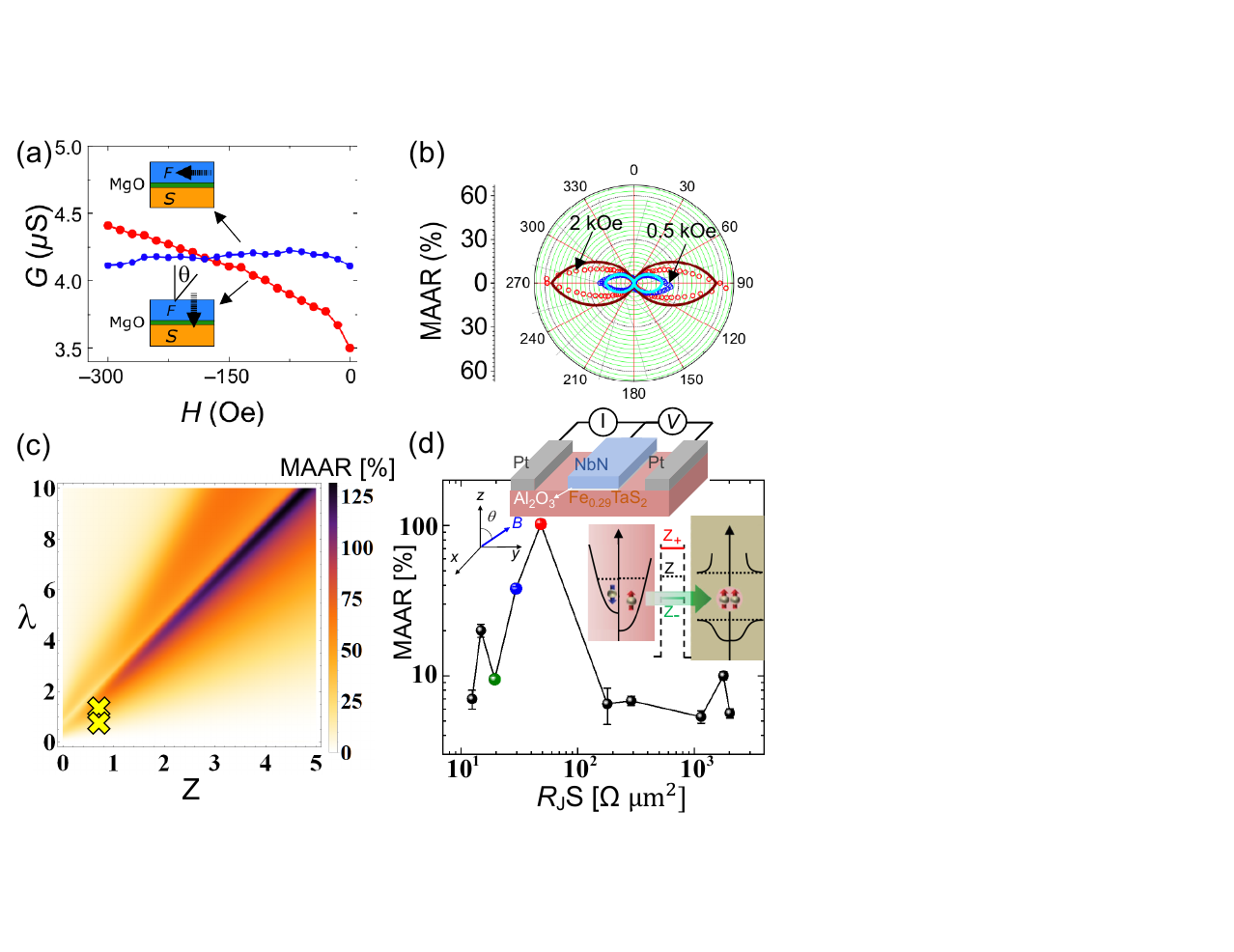}
    \vspace{-2.0cm}
    \caption{(a)   Evolution of $G(V=0)$ in Fe/MgO/V junctions, with IP and OOP $H$. The two remanent perpendicularly oriented {\bf M} (black thick arrows) with different $G(0)$ reveal MAAR 
    $\sim17~\%$  at $H = 0$. (b) OOP MAAR at $H =0.5, 2~$kOe (blue, red dots) and indicated by arrows compared with a phenomenological model (outer red and inner blue lines). The results are for $T=0.3~K$. Adapted from \onlinecite{Martinez:2020}.
    (c) OOP MAAR amplitude at $V=0$ 
    as function of the interfacial barrier, $Z$, and the 
    Rashba SOC, $\lambda$, for spin polarization $P=0.7$. Crosses:the parameters  for Fe/MgO/V junctions~\cite{Martinez:2020}.
    From \onlinecite{Vezin2020}.
    (d) OOP MAAR amplitude of Fe$_{0.29}$TaS$_2$/Al$_2$O$_3$/NbN junctions (upper inset) on the interface resistance area product, $R_JS$, near $V=0$, 
    where $R_J=V/I$. SOC-modified barrier (lower inset).
    From \onlinecite{Cai2021:NC}.
    }
    \label{fig:MAAR_MgO}
\end{figure}
Even a small $P$ and $\lambda$ yield a remarkable increase in the zero-bias conductance, due to spin-flip Andreev reflection~\cite{Vezin2020}. For $P=40\%$ the MAAR is already $10$ times greater than the normal-state TAMR. In the half-metallic limit, the MAAR depends universally on spin-orbit fields 
only~\cite{Hogl2015}.

The existence of a large OOP MAAR has been demonstrated in all-epitaxial Fe/MgO/V junctions~\cite{Martinez:2020}. By defining an angle $\theta$, measured between
{\bf M} and the interface normal, both an OOP TAMR and MAAR can
be expressed from the magnetoanisotropy of the conductance, $G$
\begin{equation}
   \rm{TAMR(MAAR)}=[G(0^o)-G(\theta)]/G(\theta).
    \label{eq:TMAAR}
\end{equation}
In the same Fe/MgO/V junction, the conductance anisotropy and TAMR can be measured by rotating {\bf M},
either by raising the temperature above $T_c$ 
or by applying the bias, to 
exceed $\Delta$ for V, as shown in Fig.~\ref{fig:MAAR_MgO}(a). While with a modest SOC in Fe/MgO/V junction
there is only a negligible TAMR of $\sim 0.01\%$ at applied field $|H|=1~$kOe, at the same temperature of 
$T=0.3$~K, a measured zero-bias conductance anisotropy in Figs.~\ref{fig:MAAR_MgO}(b)-(d) reveals that 
MAAR is enhanced by several orders of magnitude. By carefully designing the magnetic anisotropies,
two remanent states with perpendicular {\bf M} in Fe/MgO/V junctions~\cite{Martinez:2020} were obtained. 
The observed giant increase of MAAR$\sim17\%$ at $H=0$ thus excludes that and enhanced MAAR is due to  
an applied magnetic field.

A large  MAAR is connected to the 
the proximity-induced equal-spin-triplet superconductivity~\cite{Vezin2020}. 
$G$ from the spin-flip Andreev reflection dominates the same triangular region, shown in Fig.~\ref{fig:MAAR_MgO}(c) to have an enhanced MAAR~\cite{Vezin2020}. 
The nonmonotonic behavior of the spin-flip Andreev reflection arises
from the effective barrier strength~\cite{Vezin2020} $Z^\pm_{\rm{eff}} = Z \pm \lambda k_\parallel/ 2\sqrt{{k_F}{q_F}},$ where $Z^+_{\rm{eff}}$ ($Z^-_{\rm{eff}}$) is for inner (outer) Rashba bands [see Fig.~\ref{fig:omega}]. When $Z\geq 0$ and $\lambda \geq 0$, $Z^+_{\rm{eff}} \geq Z$ cannot be suppressed. However, at $k_\parallel= (2Z/\lambda)\sqrt{k_F q_F}$, $Z^-_{\rm{eff}}$ becomes 
transparent and gives a dramatically increased $G$.  
The maximum of the total $G$ 
is achieved when the amount of the open channels, $\propto k_\parallel$, is maximized. Therefore, the maximum spin-flip Andreev reflection is
near $q_F= (2Z/\lambda)\sqrt{k_F q_F}$, i.e., $\lambda=2Z$ 
for $k_F = q_F$.

Unlike common expectations that a strong SOC is desirable for equal-spin-triplet superconductivity, 
a desirable SOC strength nonmonotonically depends on the interfacial barrier.  
This trend is confirmed in
Fig.~\ref{fig:MAAR_MgO}(d) for
quasi-2D van der Waals (vdW) F/S junctions 
where, together with a large MAAR, such a measurement  
supports the equal-spin-triplet superconductivity~\cite{Cai2021:NC}. 
Conversely, while a weak interfacial barrier that enables a robust proximity-induced superconductivity seems suitable to enhance the spin-triplet superconductivity, its largest contribution is obtained for the interfacial barrier that nonmonotonically depends on the SOC strength. 
A large $G$ anisotropy is also found in all-vdW F/S tunnel junctions~\cite{Lv2018:PRB,Kang2021:X}. MAAR not only supports large spin-valve signals with
a single F, but its analysis could be used to 
identify equal-spin-triplet superconductivity and probe
elusive MZM. For MAAR, it would be important to realize
a tunable SOC in a {\em single} sample.

Combined interfacial Rashba and Dresselhaus SOC has 
been investigated in \onlinecite{Hogl2015,Costa2019,costa_prb_20}.
Since the interfacial 
barrier depends on spin and $\bm{k}$, 
it was found that so-called skewed Andreev processes, where the reflection amplitude is asymmetric in 
$\bm{k}$ space, resulted in a large anomalous Hall effect.
Another manifestation of spin-anisotropy due to SOC 
occurs at the interface of a ballistic S/HM bilayer when the symmetry breaking axis $\bm{n}$ of the spin--orbit field is rotated~\cite{Johnsen2020}. Such an effect may be achieved by combining bulk and interfacial SOC. Depending of the direction of $\bm{n}$ a modulation of was found due to anisotropic conversion of the conventional $s$-wave even frequency superconducting correlations into other pair correlations of different symmetries.

\subsection{Josephson junctions}

JJs with weak links featuring SOC have been studied for decades in the form of supercurrents through 2DEGs~\cite{Kawakami_85_2DEG,Mayer2020:NC}. Recent developments for JJs involving SOC typically include magnetic elements for
(i) inducing MZM and related topological phenomena, (ii) creating long-ranged triplet supercurrents carrying both charge and spin, or (iii) creating phase-batteries where the ground-state phase-difference in the JJ is arbitrary (not only 0 or $\pi$).

\textit{Long-ranged triplet supercurrents.} Whereas magnetically inhomogeneous structures are known to support long-ranged spin-polarized supercurrents in diffusive systems, despite the pair-breaking effect of an exchange field, SOC can accomplish this in homogeneous ferromagnets. 
Consider an inhomogeneous {\bf M} along the $x$-direction, such as a domain wall, written as $\boldsymbol{h} = h \sin(Qx)\hat{\boldsymbol{y}} + h \cos(Qx)\hat{\boldsymbol{z}},$

where $Q$ is the wavevector describing the {\bf M} rotation~\cite{bergeret_prb_14}. By performing a unitary transformation $U$ on the Green function describing such a system, corresponding to a local SU(2) rotation $U(x) = \e{-\frac{i}{2} Qx\sigma_x}$, one finds that the resulting equation of motion for the transformed Green function describes a system with homogeneous {\bf M} $\boldsymbol{h} = h\hat{\boldsymbol{z}}$, but now with an effective SOC which enters the gradient operator $\tilde{\nabla}$  like an SU(2) gauge field $\tilde{\nabla} = \nabla + \frac{iQ}{2}[\sigma^x,\cdot]\hat{\boldsymbol{x}}$ \cite{bergeret_prb_14}.
The result of this transformation is that the singlet-triplet conversion in a S/inhomogeneous F structure is equivalent to the singlet-triplet conversion in a S/homogeneous F structure with SOC. Considering a ferromagnetic nanowire, with an easy-axis anisotropy, proximitized to superconducting leads, and neglecting all forms of intrinsic SOC, one can also produce long-ranged triplet correlations by bending the nanowire. If the resulting curvature is not so large that the exchange interaction of {\bf M} overcomes the anisotropy, {\bf M} follows the bend of the wire, thereby producing a rotation of {\bf M}. As a result, an artificial, effective SOC appears.  %
This can therefore give rise to long-ranged triplet supercurrents and induce 0 to $\pi$ transitions in JJs~\cite{salamone_prb_21}. Other ways to manipulate supercurrents via SOC were discussed in \onlinecite{shekhter_prl_16, shekhter_prb_17, entin_ltp_18}.

The first experiments to detect long-ranged triplet were carried out in JJs with disordered magnetic interface~\cite{keizer_nature_06} or 
spin-mixer layers~\cite{Robinson2010,Khaire2010,Eschrig2011:PT}. These spin-mixer layers generated the triplets which were subsequently passed through a thick F (e.g. Co) to filter out the singlets. 
Therefore, the proposal to create triplets in JJs with SOC weak links is attractive as it removes the requirement of complex spin-mixer layers. 

So far, direct experimental evidence in this direction in thin-film hybrids have been inconclusive. Initial attempts~\cite{satchell_prb_18} with Nb/Pt/F/Pt/Nb, where F is a synthetic antiferromagnet composed of Co/Ru/Co showed a significant enhancement in the characteristic voltage of the JJs compared to devices without the Pt layer. However, the decay length of the supercurrent as a function of the Co layer thickness was not as expected for long-range triplets. The major limitation is the predominant IP magnetic anisotropy of the F layer instead of the canted magnetic anisotropy required to observe the long-range triplets~\cite{Jacobsen2015}. Overcoming this limitation by replacing the Pt/F/Pt weak link with a [Co/Ni]$_n$/Co multilayer with a canted magnetic anisotropy, also failed to show evidence of triplet supercurrents. It is not clear whether this discrepancy between theory and experiments is merely due to a poor singlet-to-triplet conversion efficiency in these systems or something more fundamental. 

 Lateral JJs with the current flowing in the plane of the layers are more flexible in terms of satisfying the conditions for SOC-mediated triplet generation. The original experiment observing supercurrent flow through half-metallic CrO$_{2}$ in a lateral JJ can equally be explained~\cite{bergeret_prb_14} as arising from the SOC in the contact region instead of surface magnetic inhomogeneity in CrO$_{2}$ as previously assumed. This SOC can be attributed to structural inversion asymmetry~\cite{Grioni_2007_PRL_Bi,Miron_Nature-Materials_2010}. In the lateral geometry in a disk-shaped JJ containing a Nb/Co bilayer, triplet supercurrents have been detected which are confined to the rim of the disk~\cite{Kaveh_2022_vortex}. This confinement arises from an effective SOC due to the vortex in Co.  

An additional advantage of the lateral geometry is the possibility to study the dependence of the triplet supercurrent as a function of the {\bf M} direction of F as proposed theoretically in \onlinecite{eskilt_prb_19} and later in \onlinecite{bujnowski_prb_19}. Here, a lateral JJ with SOC is in contact with an underlying F with IP anisotropy. The supercurrent was shown to be highly sensitive to the IP {\bf M} rotation with the triplet supercurrent reducing by several orders of magnitude with a $\pi/2$ rotation. This is a clear evidence of the presence of triplet supercurrents, and the device also acts as a magnetic transistor for supercurrents  without the constraint for complex magnetic anisotropies.

While the normal-state properties of heterostructures typically consider $k$-linear SOC described by models such as Eq.~(\ref{eq:Rashba}), there is a
growing class of materials where the $k$-cubic SOC is not just a small perturbation, but a dominant contribution~\cite{Winkler2002:PRB,Krich2007:PRL,Nakamura2012:PRL,Cottier2020:PRB,Liu2018:PRL}. The role of such
SOC in JJs is largely unexplored~\cite{Alidoust2021:PRB}. The corresponding Hamiltonian can be written as $H_\mathrm{SO} = (i\alpha_c/2)(k_-^3\sigma_+ - k_+^3\sigma_-)-
(\beta_c/2)(k_-^2k_+\sigma_+  +  k_+^2k_-\sigma_-),$ using cubic strengths $\alpha_c$ and $\beta_c$, for  Rashba and Dresselhaus terms, 
where $k_\pm = k_x \pm i k_y$, and $\sigma_\pm = \sigma_x \pm i \sigma_y$. 
Without symmetry constraints to prevent $k$-linear SOC, the relative
$k$-cubic SOC contribution depends on the carrier density~\cite{Zutic2004:RMP,Krich2007:PRL}.
Therefore, assuming a simple $k$-linear SOC may not in general be adequate. HMs in superconducting heterostructures do not have small Fermi pockets and the interpretation for the long-range triplet decay may need to be revisited.
Furthermore, from Fig.~2, we see that the emergent interfacial SOC 
is not $k$-linear and can strongly modify transport properties.

A hallmark of JJs with cubic SOC goes beyond current-phase relations (including
the anomalous Josephson effect, discussed below in the context of phase batteries) 
and also influences the spin structure and symmetry properties of superconducting proximity effects. 
Unlike the $p$-wave for linear SOC, the $f$-wave symmetry of superconducting correlations is the fingerprint for cubic SOC, which supports MZM~\cite{Alidoust2021:PRB}. Cubic Rashba SOC also provides an effective low-energy description for the heavy holes in Ge-based planar JJs~\cite{Luethi2022:X,Tosato2022:X}.

\textit{Phase-batteries.} The supercurrent flowing through JJs depends sensitively on the phase difference $\phi$ between the superconductors. A finite phase difference usually drives a supercurrent through the system and the ground-state of the system is usually $\phi=0$ or $\phi=\pi$. But this is not always the case \cite{golubov_rmp_04}. 
The phase difference $\phi$ of the superconducting order parameters by $2\pi n$ where $n$ is an integer should correspond to exactly the same physical state. Moreover, a dc supercurrent flows if a gradient exists in the phase of the superconducting order parameter in the junction. If $I(0)=0$, it follows that $I(2\pi n)=0$. Finally, performing a time-reversal operation on the system must reverse any supercurrent present. Since time-reversal includes complex conjugation, the phase changes sign and $\phi \to -\phi$. Therefore, one usually has $I(\phi) = -I(-\phi)$. 

From the above properties, it follows that $I=0$ whenever $\phi=\pi n$. Therefore, the ground-state phase difference of a JJ, being the state with no supercurrent, is usually either 0 or $\pi$. This can change if time-reversal symmetry (TRS) and inversion symmetry is broken. In JJs with superconductors breaking TRS, such as $d+\ii d$ superconductors, the relation $I(\phi) = -I(-\phi)$ is not necessarily fulfilled~\cite{Liu_srep_2017}. Instead, the phase difference $\phi$ that minimizes the Josephson energy of the system can be neither $0$ or $\pi$, but a different value denoted $\phi_0$. There is no supercurrent for the ground-state phase difference $\phi_0$, so that $I(\phi_0) = 0$. 

JJs where the ground-state phase difference is neither 0 or $\pi$, but some arbitrary value $\phi_0$ are known as $\phi_0$ junctions~\cite{buzdin_prl_08}. This behavior is also found
in earlier studies by~\onlinecite{Geshkenbein1986:JETPL,Millis1988:PRB},
considering JJ with unconventional superconductors, SOC, and magnetically-active
interfaces.
Assuming that the value of $\phi_0$ is tunable, a suitable name for such systems is in fact \textit{phase-batteries}. By tuning $\phi_0$ via external parameters,  one provides a phase bias to a macroscopic wavefunction in a quantum circuit. This is conceptually similar to how a classical battery provides a voltage bias. The question is then if controllable $\phi_0$ junctions can 
be tailored by combining materials with the right properties into a JJ.

One way to achieve a phase-battery using conventional superconductors is combining antisymmetric SOC with a spin-splitting Zeeman field in the weak link. This breaks TRS and inversion symmetry which can result in a finite supercurrent even at zero phase difference~\cite{zazunov_prl_09, buzdin_prl_08}. Here "breaking inversion symmetry" means that some particular operation on the spatial degrees of freedom in the system, such as a mirror, parity, or rotation operation (or combination thereof) does not leave the Hamiltonian of the weak link invariant. The precise description of which spatial symmetry that needs to be broken for the $\phi_0$ junction to appear is system-specific, depending \eg on the direction of the spin-splitting field~\cite{liu_prb_10, rasmussen_prb_16}.

A microscopic explanation of the $\phi_0$-effect in JJs with quantum dots (QDs) was given in~\onlinecite{zazunov_prl_09, szombati_natphys_16}. The Hamiltonian of a two-level QD with SOC and spin-splitting is~\cite{szombati_natphys_16} $H_\text{QD} = (E_\text{orb}\tau_z - \mu\tau_0) \sigma_0 + B \tau_0\sigma_z +\alpha\tau_y\sigma z,$
where $\mu$ is the chemical potential, $E_\text{orb}$ is the orbital energy, $\alpha$ parametrizes the SOC strength, $B$ is the Zeeman splitting, while $\tau_{0,x,y,z}$ and $\sigma_{0,x,y,z}$ are the identity and Pauli matrices acting on orbital and spin space, respectively. Without SOC, $\alpha=0$ and the two orbitals do not mix. Transfer of electrons in the Cooper pair through the QD then takes place in one level at the time. Consider one electron tunneling via level 1 and the second via level 2: the corresponding matrix element for such a process is $t_L^{(1)} t_R^{(1)} t_L^{(2)} t_R^{(2)},$ where $t_L^{(i)}$ are the hybridization amplitudes between level  $i$ in the QD and the left lead and can be taken as real for $\phi=0$. Similarly, for $t_R{(i)}$. 
Therefore, the matrix element describing tunneling from right to left is exactly the same as left to right when $\phi=0$, 
hence no current flows.

When $\alpha\neq 0$, the $H_\text{QD}$ eigenstates %
are a mix of the two orbital states. As shown in \onlinecite{szombati_natphys_16}, this results in new single-level hybridization amplitudes $T_L^{(1,2)}$ for levels 1, 2 with the left lead (determining the probability for electron transfer between the level and the lead). For spin-$\uparrow$ electrons, %
$T_L^{(1)} = t_L^{(1)} \cos\epsilon + \ii\sin\epsilon t_L^{(2)}$, $T_L^{(2)} = t_L^{(2)} \cos\epsilon -\ii\sin\epsilon t_L^{(1)}.$
The expressions for $T_R^{(i)}$ are obtained by $L \to R$. For spin-down electrons, $\pm$ signs above are exchanged. 

A key observation is that the amplitudes $T_{L(R)}^i$ are now complex. This means that as electrons make their way across the QD, they gain a finite phase. The phase of the resulting matrix element is opposite for electrons tunneling in one direction (say, left to right) compared to the opposite direction (right to left). Since the imaginary part of the the rightward and leftward total tunneling coefficients are then different, leftward and rightward tunneling do not cancel each other exactly for a given spin species $\sigma$. If the tunneling probabilities are now also different \textit{in magnitude} for spin $\uparrow$ and $\downarrow$ (for $B \neq 0$), the Cooper pairs acquire a net phase upon tunneling despite $\phi=0$.

$\phi_0$ has also been predicted in JJs with metallic interlayers, such as multilayered ferromagnets~\cite{braude_prl_07, liu_prb_10, grein_prl_09, kulagina_prb_14} and through metallic weak links that contain both SOC and ferromagnetic order~\cite{buzdin_prl_08}. This effect can be understood in terms of Andreev bound states, comprised of a counterpropagating electrons and holes which transfer Cooper pairs between superconductors via Andreev reflection. These bound states come in pairs $\pm E_i$ where $i$ is an index characterizing internal degrees of freedom such as the spin of the electron and hole that comprise the bound state. Consider first a simple S/F/S JJ. In the limit of weak Zeeman splitting $h$ and assuming a high-transparency junction for simplicity, one finds energies~\cite{annunziata_prb_10} $
E_i = E_\sigma = E_0\cos(\phi/2 + \sigma ch),\; \sigma=\pm 1,$
where $c$ is a geometry-dependent constant  and $E_0=\Delta$. The current carried by these Andreev bound states is proportional to $dE_\sigma/d\phi$. While each bound state is phase-shifted by $\sigma ch$, the total current at $\phi=0$ vanishes since the magnitude of each current is identical. 

The situation changes in a S/F$_1$/F$_2$/F$_3$/S junction. When the magnetization ${\bf M}_i$ of the ferromagnets is such that the spin-chirality defined as
${\bf M}_1 \cdot ({\bf M}_2 \times {\bf M}_3)$
is non-zero, an anomalous $\phi_0$ JJ emerges. When all {\bf M} are perpendicular to each other, the spin-chirality is maximized. The reason for why the spin-chirality needs to be finite is precisely that both time-reversal symmetry and inversion symmetry are now broken in a manner that permits the $\phi_0$-effect (recall that ${\bf M}_j$ is a pseudovector). The Andreev bound states are~\cite{liu_prb_10} $E_i = E_\eta = E_{\eta,0} \cos(\phi + \eta c'h),\; \eta=\pm 1,$ where $c'$ is a new constant, under the simplifying assumptions that the Zeeman-splitting in each F layer is equal and that the spin-chirality product is maximized. The index $\eta$ is related to the spin of the Andreev bound state. The crucial difference from the S/F/S case is that the amplitude $E_{\eta,0}$ of the bound state is now unequal for the two bound states $\eta=\pm1$. Therefore, the total current at $\phi=0$ does not cancel out and a net anomalous supercurrent exists even at zero  phase difference.

\begin{figure}[t]
    \centering
    \includegraphics[width=\columnwidth]{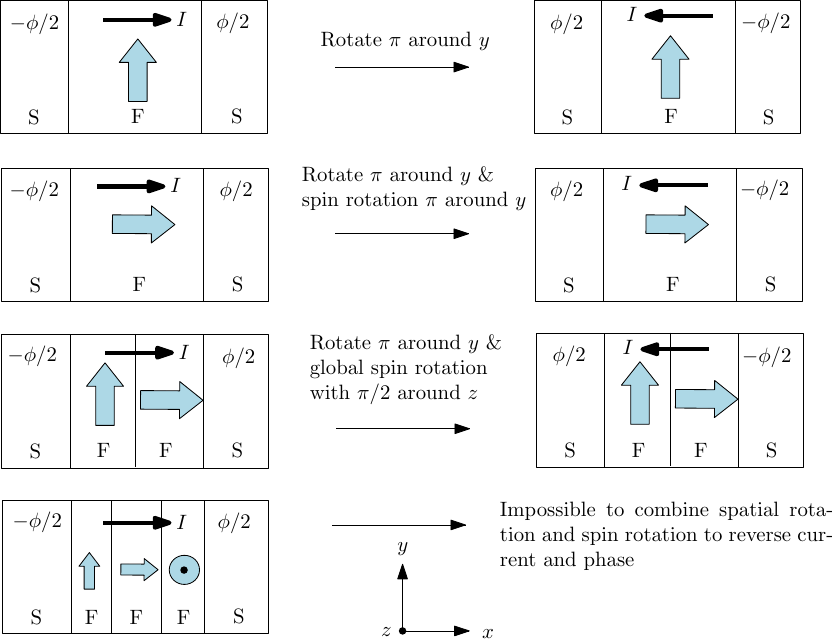}
    \caption{
    Illustration of how a combination of spatial rotation of the entire junction and a global spin rotation allows one to prove $I(\phi) = -I(-\phi)$ for magnetic JJs with zero spin-chirality. The blue (light gray) arrows show the {\bf M} in each layer.}
    \label{fig:phi01}
\end{figure}

A pictorial argument shows when a $\phi_0$-effect can appear, which is intuitively easier to understand than using matrix symmetry operations applied to the Hamiltonian of the system. Consider first a magnetic JJ with an arbitrary number of magnetic layers. Without SOC, there is no coupling between the spin degree of freedom and orbital motions of the electrons. Therefore, a global spin rotation should leave the supercurrent invariant: if all {\bf M} are rotated in the same way, the current stays the same. The goal is now to use this global spin rotation invariance as well as a spatial rotation of the entire JJ  to prove that $I(\phi) = -I(-\phi)$. As explained in Fig.~\ref{fig:phi01}, this is possible to accomplish for arbitrary {\bf M} directions with both one and two ferromagnets, but not with three (lower row) if the spin-chirality is finite. The pictorial proof shown in the figures can thus be used to prove if the $\phi_0$-effect is absent. We note that the same type of pictorial proof should be possible to use for non-reciprocal dissipative transport by replacing the superconducting phase differences $\pm\phi/2$ with voltages $\pm V/2$. 

\begin{figure}[t]
    \centering
    \includegraphics[width=\columnwidth]{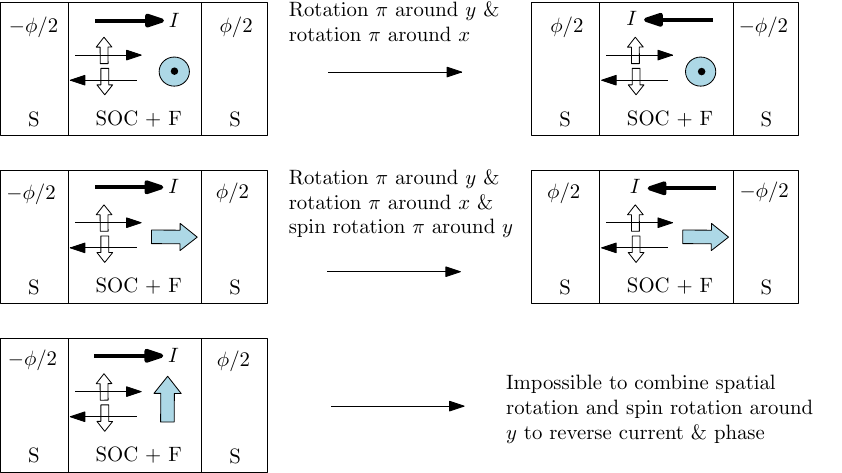}
    \caption{Illustration of how a combination of spatial rotation of the entire junction and a particular global spin rotation allows one to prove $I(\phi) = -I(-\phi)$ for magnetic JJs with SOC. A Rashba-type SOC $k_x\sigma_z$ is considered, and the illustration of SOC in the figure indicates that $\uparrow$ have lower energy when moving in one direction whereas $\downarrow$ have lower energy when moving in the opposite direction. }
    \label{fig:phi02}
\end{figure}

We can use the pictorial proof also for a S/F/S junction with SOC to infer which {\bf M} directions that do not permit a $\phi_0$-state. We consider a Rashba-type SOC $\propto \alpha \sigma_z k_x$ in a 1D geometry for simplicity since this suffices to show the principle. This is shown in Fig. \ref{fig:phi02}. The upper row shows that if {\bf M} points in the $z$-direction, physically rotating the entire junction two times brings it back to its original state except for a reversed current and phase. Therefore, one concludes $I(\phi) = -I(-\phi)$: no $\phi_0$-state. When the {\bf M} points along the $x$-direction, a global spin rotation around the $y$-axis is still permitted without changing the supercurrent. The reason is that performing this spin rotation does not change the SOC-term nor the absolute or relative magnitude of the momentum-dependent total exchange-field of the carriers. Combined with a physical rotation of the entire system, one proves again that $I(\phi) = -I(-\phi)$. This is not possible to do when {\bf M} points in the $y$-direction, consistent with the known result that such a system hosts a $\phi_0$-state. 
A further manipulation of such $\phi_0$-state is possible with the contribution of Rashba and Dresselhaus SOC~\cite{Alidoust2020:PRB} and experimentally-demonstrated gate-controlled 
SOC~\cite{Mayer2020:NC,Dartiailh2021:PRL}.

Besides the phase-shift obtained due to the broken time-reversal and inversion symmetry in the junction, the magnitude of the critical current also becomes direction-dependent, which we discuss in more detail in the following subsection. Finally, the presence of SOC in magnetic JJs has also been shown to induce electrically-controlled {\bf M} dynamics~\cite{nashaat_prb_19} and specific magnetization
trajectories along IV-characteristics of $\phi_0$ JJs \cite{shukrinov_physusp_22}, and anomalous Gilbert damping and Duffing features~\cite{shukrinov_prb_21, abdelmoneim_prb_22}.

\subsection{Supercurrent diodes}

There has been a resurgence in the interest of the superconducting diode effect,
which shares a curious
history with the spin Hall effect, predicted decades 
before~\cite{Dyakonov1971:PL,Dyakonov1971:JETPL} 
the current terminology was established~\cite{Hirsch1999:PRL}.
\onlinecite{Swartz1967:PR} found a rectifying behavior in a superconducting Pb-based bimetallic strip in an applied magnetic field. The magnitude of $I_c^+$ in the forward direction mismatched to $I_c^-$ in the reverse direction, the behavior noted independently in
the abstract of \onlinecite{Edelstein1996:JPCM}. 
This means that there exists a magnitude range $I_c^- < I < I_c^+$ where the current $I$ is dissipationless in one direction, but resistive in the other.
The term ``Josephson diode” was %
used by~\onlinecite{Hu2007:PRL} with the proposed
implementation of the $p$- and $n$-doped region resembling conventional semiconductor 
diodes without SOC, but with a broken inversion symmetry and a rectifying behavior~\cite{Shockley1949:BSTJ}. Before the recent observation by~\onlinecite{ando_nature_20},
early experimental~\cite{Touitou2004:APL,Vodolazov2005:PRB} and theoretical~\cite{grein_prl_09,Silaev2014:JPCM} studies of the superconducting diode effect with
ferromagnets, that do not invoke SOC, were summarized in~\onlinecite{Satchell2023:JAP}.

Many recent experimental realizations
of the superconducting diode effect closely follow theory~\cite{Reynoso2008:PRL}, appearing in JJs with spin splitting and SOC. However, while some measurements show the diode effect~\cite{Mayer2020:NC, Dartiailh2021:PRL}  
prior to ~\onlinecite{ando_nature_20}, such an effect was overlooked. Figures~\ref{fig:MZM_JJ}(c), (d) show such a supercurrent diode effect, where the normalized critical current asymmetry reaches $\approx10-20\,$\%~\cite{Dartiailh2021:PRL}.

Theory~\cite{davydova_sciadv_22, scammell_2d_22, yuan_pnas_22, he_njp_22, ilic_prl_22, hou_arxiv_22} and experiments~\cite{ando_nature_20, Pal_2022_NatPhys,Baumgartner2022:NN} have investigated non-reciprocal critical currents in superconducting wires or films. The supercurrent diode effect in a JJ is related to the appearance of an anomalous phase, but it is possible to have a $\phi_0$ JJ 
without any accompanying diode effect. A current-phase relation of $I=I_0\sin(\phi+\phi_0)$ gives an anomalous phase difference, but no diode effect since the positive and negative critical currents match. Since $\delta \sin(\phi+\phi_0)$ can be written as $\alpha\sin\phi + \beta\cos\phi$ for real coefficients $\alpha,\beta,\delta,\phi_0$, 
for the diode effect one additionally requires higher-order harmonics in the current-phase relation beyond $\sin\phi$ and $\cos\phi$, in order to achieve different magnitudes of the positive and negative critical current~\cite{Baumgartner2022:NN}. A skewed current-phase relation is therefore a necessary condition.

\begin{figure}%
    \centering
\subfloat{{\includegraphics[width=3.7cm]{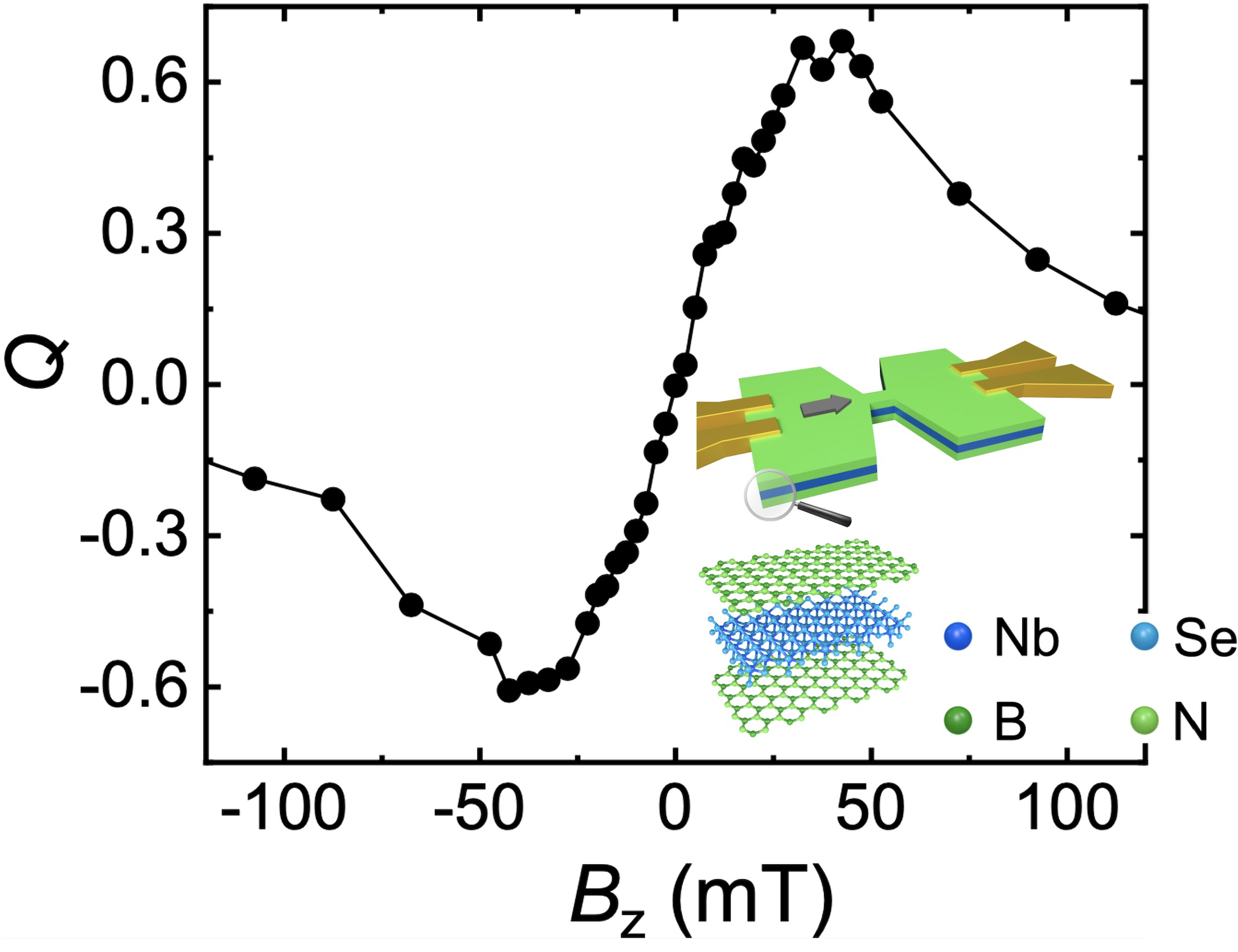} }}%
    \qquad
\subfloat{{\includegraphics[width=4cm]{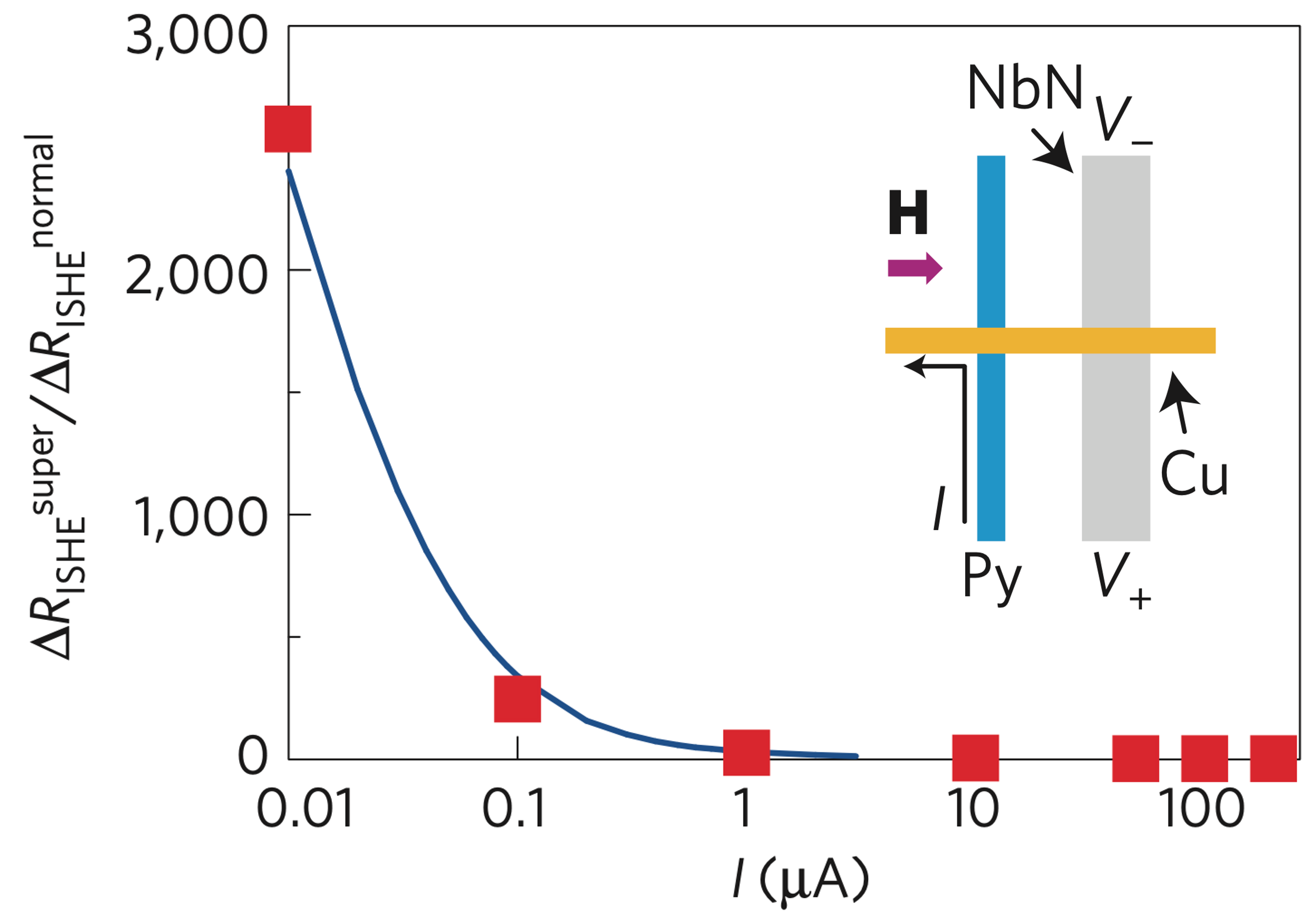} }}%
    \caption{\textit{Left}: The supercurrent rectification efficiency, $Q \equiv2({I_{c}^{+}-\lvert{I_{c}^{-}\rvert}})/({I_{c}^{+}+\lvert{I_{c}^{-}\rvert}})$, as a function of the OOP applied magnetic field measured at 1.3 K. $Q$ is maximum around 35 mT. Inset: the device structure with a 250 nm long and wide central constriction and the $z$ direction is perpendicular to the crystal plane. 10 nm hBN encapsulates 2, 3 or 5-layer NbSe$_2$. 
    From \onlinecite{Bauriedl_2022_NatComm}. \textit{Right}: Inverse spin Hall signal at 3 K, quantified by $\Delta R_{\mathrm{ISHE}}^{\mathrm{super}}$ versus injected spin current $I$. The signal is normalized by the measured value at 20 K in the normal state $\Delta R_{\mathrm{ISHE}}^{\mathrm{normal}}$. The blue (light gray) line is the fit to the model as discussed in~\onlinecite{wakamura_natmat_15}.}%
    \label{fig:diode1}%
\end{figure}

Several studies have focused on superconducting systems where time-reversal and inversion symmetry breaking are modelled via ferromagnetism/a magnetic field and antisymmetric SOC interactions, such as Rashba SOC. Nonreciprocal supercurrents have also been observed in materials with valley-Zeeman SOC where, unlike the Rashba SOC, the rectification of supercurrent depends on the OOP magnetic field~\cite{Bauriedl_2022_NatComm}. Figure~\ref{fig:diode1} shows rectification efficiency of 60$\%$ measured in transition metal dichalcogenide NbSe$_2$ sandwiched between hBN layers which is significantly larger than those observed in Rashba SOC systems. The rectification saturates at low temperature with a maximum observed around $T=T_c/2$, unlike the observations by \onlinecite{ando_nature_20} where the diode effect was observed near $T_c$. The unusual temperature dependence together with the rectification appearing with an OOP applied magnetic field indicates a fundamentally different origin of the diode effect in comparison to those observed in Rashba superconductors.

There are aspects of the supercurrent diode effect which remain poorly understood. For instance, the experimental observation of the diode effect in \onlinecite{ando_nature_20} only occurred near $T_c$, vanishing far below $T_c$. This suggests that a key role is played by fluctuations in $\Delta$.

\subsection{Spin-pumping}

Traditional studies of spin transport 
involve quasiparticle injection at voltages above the superconducting gap. They show evidence for spin and charge decoupling~\cite{Charis_2013_N.Phys,Hubler_12} and enhancement of the spin relaxation times ~\cite{Parkin_2010_NM, wakamura_prl_14}. Previous experiments demonstrated that Andreev reflection excludes transport of dynamically-driven spin currents through the superconducting energy gap and so the spin-current-induced broadening of the FMR linewidth is suppressed by the opening of the superconducting gap~\cite{Bell_2007_PRL}.

\onlinecite{Jeon2018} compared FMR results on Nb/Py/Nb trilayers with Pt/Nb/Py/Nb/Pt structures in which the outer Pt layers are effective spin sinks with strong SOC. The authors investigated the $T$-dependence of the FMR linewidth ($\mu_{0}\Delta{H} \propto \alpha$) and the resonance field $\mu_{0}H_\mathrm{{res}}$ across $T_{C}$. Where Pt (or other large SOC spin sinks) are present, a substantially increased FMR damping for an S layer thickness of the order the coherence length is interpreted as evidence for superconducting pure spin (triplet) supercurrent pumping. The key mechanism driving the spin current through superconducting Nb involves an interaction of the SOC in Pt with a proximity exchange field from Py, which passes through Nb. Theoretically, this requires Landau Fermi liquid interactions and a non-negligible spin splitting in Pt, creating a triplet channel in the superconducting density of states of Nb around zero energy~\cite{PhysRevB.98.104513}. \onlinecite{Jeon:2020} substituted Pt for a perpendicularly magnetized Pt/Co/Pt spin sink allowing tunability of the pure spin supercurrent by controlling the {\bf M} angle of Co with respect to the SOC.

An alternative explanation to the FMR results were suggested by \cite{Muller:2021} where they studied spin injection into superconducting NbN from an adjacent Py layer. Here, the enhanced $\mu_{0}\Delta{H}$ is attributed to an increased inhomogeneous broadening rather than increased damping. It is interesting to note that even though \cite{Muller:2021} uses thicker NbN compared to the coherence length, they see an enhanced $\mu_{0}\Delta{H}$ while \cite{Jeon2018} only observes this enhancement for thinner Nb layers. For thicker Nb layers, they observe a decrease of the linewidth below the $T_{C}$ indicating that comparison of results between different systems may not be straightforward and more work is required to understand the discrepancies. 

\subsection{Spin-Hall phenomena with superconductors}

The spin Hall effect~\cite{sinova_rmp_15} and its inverse are key in spintronics, providing a method to electrically detect spin currents. The spin Hall effect also takes place theoretically in superconducting materials where a longitudinal flow of charge or spin converts to a transverse flow of spin or charge. Experimentally demonstrating a superconducting spin Hall effect would provide a means to electrically detect the polarization of spin supercurrents and potentially control spin in the superconducting state.

To understand this prospect, we start with the magnetoelectric phenomena in superconductors which were already studied in~\onlinecite{edelstein_prl_95}. Considering S lacking spatial reflection symmetry, Edelstein predicted that the supercurrent must be accompanied by an induced spin-polarization among the itinerant electrons. Spin Hall effects in superconductors were later considered in~\onlinecite{malshukov_prb_08}, predicting an induced edge spin polarization in a JJ with a spin-orbit coupled layer separating the superconductors. \onlinecite{kontani_prl_09} considered instead the dissipative spin Hall effect in S with Rashba SOC,  
predicting a large negative spin Hall conductivity in the superconducting state. Several works followed, considering the spin Hall effect in different 
JJ geometries, including 
an AC Josephson bias~\cite{malshukov_prb_10, malshukov_prb_11}.

An important experimental breakthrough of \onlinecite{wakamura_natmat_15}, motivated by
theoretical predictions~\cite{Takahashi2002:PRL,Takahashi2008:JPSJ,Maekawa:2006}, reported a giant quasiparticle-mediated inverse spin Hall effect in the superconductor NbN, which exceeded the effect in the normal state by three orders of magnitude. The signal diminished when the distance between the voltage probes in the setup exceeded the charge imbalance length, indicating that the inverse spin Hall signal was indeed carried by quasiparticles. This quasiparticle-mediated spin Hall effect in the superconducting state was measured by the spin
absorption technique using a lateral structure composed of Py and a superconducting NbN wire joined by a nonmagnetic Cu bridge as illustrated in the inset of Fig.~\ref{fig:diode1}. The spin current injected via Py diffuses towards the NbN wire and is partly absorbed by it owing to the high SOC in NbN where it is converted to a charge current (quasiparticle current in the superconducting state) via the inverse spin Hall effect.

Theoretical studies followed shortly~\cite{espedal_prb_17, huang_prb_18}, utilizing the quasiclassical theory of superconductivity. With this methodology, one derives kinetic equations for the distribution functions for energy, charge, and spin-excited modes in the system, which permits computation of currents. The coefficients in these kinetic equations are determined by the spectral properties of the material and therefore can be very different in the normal and superconducting state. \onlinecite{espedal_prb_17} computed the various contributions to the spin Hall effect in a conventional superconductor, including side-jump, skew scattering, and anomalous velocity operators. They found that the inverse spin Hall current (i.e. a charge current) $j^\text{sH}_i$ flowing in the $i$-direction could be computed from the injected spin-current $j^s_{jk}$ flowing in the $j$-direction and polarized along the $k$-direction according to
\begin{align}
    j^\text{sH}_i = \theta^\text{sH}\epsilon_{ijk} j^s_{jk}
\end{align}
where $\epsilon_{ijk}$ is the Levi-Civita tensor while $\theta^\text{sH}$ is the spin Hall angle. In the superconducting state, 
in the diffusive limit, it is $\theta^\text{sH} = \chi^\text{sH} N_S D /D_E.$
Here, $\chi^\text{sH}$ is the normal-state spin Hall angle whereas $D$ and $D_E$ are renormalized, energy-dependent diffusion coefficients in the superconducting state with a ratio $D/D_E$ which depends only weakly on energy. Instead, the key factor in the expression for $\theta^\text{sH}$ is the superconducting density of states $N_S$ which depends strongly on energy near the gap edge $E \simeq \Delta$. This result predicts a strong enhancement in the spin Hall angle for quasiparticle energies near the gap $\Delta$, consistent with the experimental findings of~\cite{wakamura_natmat_15}. However, the total accumulation due to the spin Hall effect, obtained by integrating over all energies, does not show nearly the same magnitude of the enhancement as in the experiment. Thus, a complete explanation of the experimental results remains an open question. Additional studies have reported efficient spin-charge conversion in spin-split \cite{jeon_acsnano_20} and Ising superconductors \cite{jeon_acsnano_21}, and a recent theoretical work \cite{kamra_arxiv_23} reports the same enhancement by orders of magnitude in the spin-split case as seen experimentally. Moreover, a theory incorporating both the role of intrinsic and extrinsic SOC in the kinetic equations determining the spin-charge conversion has been reported \cite{virtanen_prb_22, Virtanen2021}.

Another interesting aspect regarding spin Hall phenomena 
involves pure supercurrent transport: conversion of Cooper pair charge currents to Cooper pair spin currents and vice versa. \onlinecite{yang_cpb_12, linder_prb_17} considered this phenomenon in JJs with interfacial SOC and a Zeeman field. Here, it was found that applying a phase gradient between the superconductors produces a transversal spin current at the junction interfaces. The physical origin of this effect stems from the $p$-wave superconducting correlations induced by the SOC. As was explained in Sec.~\ref{sec:SCh}, they can give rise to equilibrium spin currents provided they are not phase shifted by $\pi/2$ with respect to the singlet correlations. In a S/F bilayer, the Zeeman field in F will indeed cause such $\pi/2$-shifted $p$-wave triplets to appear, having seemingly no observable consequences. However, in a JJ with a phase gradient, $p$-wave triplets induced at the interface to one of the superconductors can be non-zero at the other S interface, modifying the singlet-triplet phase-shift, and thereby yielding spin currents. 

In typical spin Hall phenomenology, an injected current in a given direction is deflected transversely to the injected current while being polarized in a direction transverse to both the injection and deflection axis. The spin Hall supercurrent considered above behaves similarly, with the polarization of the triplet Cooper pairs carrying the current taking the role of the spin-polarization of the resistive current in the conventional case. The superspin Hall effect was later studied in a finite size structure~\cite{risinggard_prb_19}, demonstrating that the induced transverse spin supercurrent flow would produce an edge spin {\bf M}, similarly to the conventional resistive spin Hall effect. The prediction of an anomalous supercurrent Hall effect in F with a nontrivial spin texture~\cite{yokoyama_prb_15} suggests that similar effects could take place in homogeneously magnetized systems with SOC, since conversion processes between singlet and triplet Cooper pairs can be shown to be similar in systems with inhomogeneous {\bf M} and systems with homogeneous {\bf M} and SOC~\cite{bergeret_prb_14}. Indeed,~\onlinecite{costa_prb_20} predicted an anomalous Josephson Hall effect and transverse spin supercurrents in JJs with a macrospin F and interfacial SOC.

Whereas the above studies considered intrinsic Rashba SOC~\onlinecite{bergeret_prb_16} predicted that dissipationless magnetoelectric phenomena like spin Hall effects with supercurrents should also occur with extrinsic (impurity) SOC. In the diffusive transport, they predicted a nondissipative spin-galvanic effect, corresponding to generation of a supercurrent by a spin-splitting field, in addition to its inverse -- a magnetic moment induced by a supercurrent. Edge {\bf M} induced by supercurrents due to interfacial SOC was studied in~\onlinecite{silaev_prb_20, linder_prb_22}. Long-ranged triplets and controllable $0-\pi$ switching has also been predicted in such systems~\cite{Mazanik2022}. Another path to achieve spin Hall-like phenomena is the exploit the orbital degrees of freedom \cite{chirolli_prl_22, mercaldo_prb_22}.

\section{Open questions and future directions}

As a growing class of novel materials is considered, the understanding of SOC in superconducting hybrid structures and how it can be utilized continues to evolve. SOC, beyond common Rashba and Dresselhaus models, modifies both Cooper pairs and quasiparticle excitations. However, in junctions with common elemental superconductors and conventional semiconductors, even simple SOC models
can accurately capture the observed generation of spin-triplet Cooper pairs, the transition to topological superconductivity~\cite{Dartiailh2021:PRL}, an anomalous phase in Josephson junctions (Mayer et al., 2020), and nonreciprocal phenomena~\cite{Nadeem2023:P}.

Compared to the generation of spin-triplet superconductivity with multiple ferromagnets or magnetic textures, the presence of SOC can simplify the structure and offer a desirable gate-controlled tunability. With a proximity-induced triplet superconductivity this gate-tunable SOC is predicted to control superconducting $T_c$~\cite{ouassou_scirep_16} and its feasibility is experimentally supported by the singlet-to-triplet transition in measurements from Figs. 5(c) and (d).  An alternative approach is to use tunable magnetic textures and the resulting fringing fields, experimentally shown to yield synthetic SOC in proximity-induced superconductivity~\cite{Desjardins2019:NM}. An open experimental challenge is to electrically control tunable magnetic configurations, employed in 
spintronics~\cite{Tsymbal:2019} which, in the context of superconducting hybrid structures, could generate spin supercurrents as well as define topological structures, not by epitaxy, but by tunable magnetic textures~\cite{Fatin2016:PRL,Zhou2019:PRB,Gungordu2022:JAP}. 

The prospect of equal-spin-triplet superconductivity poses many questions for dynamical and nonequilibrium phenomena. For example, how could the resulting spin supercurrents modify magnetization dynamics?  In the absence of SOC, supercurrent-induced magnetization dynamics was proposed two decades ago~\cite{waintal_prb_02} and studied from first principles~\cite{Wang2010:PRB}. The role of SOC, including the spin-orbit and spin-transfer torques generated by spin-triplet supercurrents~\cite{zhao_prb_08}, was also 
considered~\cite{Brydon2011:PRB,hals_prb_16,nashaat_prb_19,Takashima2017:PRB}. Experimentally, supercurrent densities are generally too low to compete with the magnetic anisotropy of ferromagnets, although they may influence a nanomagnet~\cite{Cai2010:PRB}. This suggests that spin supercurrents could modify various magnetic 
nanotextures~\cite{buzdin_prl_08, linder_prb_11}, including domain walls and skyrmions~\cite{kulagina_prb_14, Tsymbal:2019, rabinovich_prl_19}, potentially to change their topology or implement superconducting counterparts of magnetic racetrack~\cite{Hayashi:Science}. 

Magnetization dynamics in superconducting hybrid structures is an example of nonequilibrium phenomena where, in addition to solving the kinetic equations, suitable boundary conditions describing the interfaces between the different layers are also required. Only recently~\cite{bobkova_prl_21} the time-dependent interplay between SOC and superconducting order in a single layer has been successfully modelled. A gate-controlled time-dependent SOC can strongly modify the current-phase relations and drive the JJ dynamics even without any bias current~\cite{Alidoust2022:PRA}. This work, supported by the experiments in planar JJs~\cite{Dartiailh2021:PRL}, has implications for superconducting spintronics, Majorana states, emerging qubits~\cite{Krantz2019:APR}, and enhanced neuromorphic computing~\cite{Crotty2010:PRE}. Changing the current-phase relations for improved $\pi$-qubits using ferromagnetic or $d$-wave 
JJs~\cite{Ioffe1999:N,Yamashita2005:PRL}, can be generalized by time-dependent gate-tunable SOC and to also realize other paths towards fault-tolerant operations~\cite{Brooks2103:PRA,simoni_acs_21,Larsen2020:PRL,mercaldo_arxiv_23,Pita-Vidal2020:PRA}.

Developing new and identifying optimal materials is an active area of investigation that will determine the progress in several challenges outlined above. 
A prominent example is the growing family of vdW materials where strong SOC can coexist with superconductivity and 
ferromagnetism~\cite{delaBarrera2018:NC,Saito2016:NP,Sierra2021:NN,Zutic2019:MT},
offering a fertile playground to realize nonreciprocal transport, anomalous Josephson effect,
topological superconductivity, and triplet supercurrents~\cite{Bauriedl_2022_NatComm,Cai2021:NC,Kezilebieke2020:N,Hu2023,Holleis2023:X,Xie2023:PRL}.  %
Controlling superconductivity in vdW materials can be enhanced with chiral 
molecules~\cite{Wan2023:X}. Their chirality-induced spin selectivity~\cite{Naaman2020:ACR} connects the chiral structure and the electron spin, which mimics the effects of SOC and magnetism and,
therefore, could support spin-triplet and topological states with conventional superconductors~\cite{Alpern2021:PRM,Ozeri2023:AMI}. %

Existing materials will 
continue to play a crucial role for future development and considerable amount of work still needs to be done to understand the full impact of interfaces, and particularly interfacial electronic and spin states. Understanding them are especially important in oxide systems where generation of triplet Cooper pairs have long been attributed to noncollinear spins or SOC at the S/F interface without direct experimental observation~\cite{Sefrioui2003,Pena2004,keizer_nature_06,Dybko2009,Kalcheim2011,Visani2012,SanchezManzano2022,Yates2013,bergeret_prb_14}. While a detailed review of the progress in oxide materials is beyond the scope of this Colloquium, we refer the reader to 
papers cited in \onlinecite{Cuoco2022} %
which could be relevant for future work on  
triplet pairing 
from SOC in oxide systems.

Finally, an interesting direction is combining the emerging field of orbitronics with 
superconductivity~\cite{fukaya_prb_18,fukaya_npj_22,mercaldo_prapp_20,mercaldo_prb_22} Since the orbital $M$ can be much greater than spin $M$, this can lead to enhanced magnetoelectric phenomena such as a large supercurrent-induced Edelstein effect~\cite{chirolli_prl_22}.

\acknowledgments
The authors thank J. A. Ouassou, E. H. Fyhn, S. Jacobsen, L. G. Johnsen, L. A. B. Olde Olthof, F. Aliev, J. Shabani, T. Zhou, W. Han, J. E. Han, A. Matos-Abiague, C. Shen, J. Fabian, R. de Sousa, M. Alidoust, D. Caso, C. Gonzalez-Ruano, V. Risingg{\aa}rd, I. Bobokova, A. Bobkov, and F. Giazotto for useful discussions. J.L. was supported by the Research Council of
Norway through Grant No. 323766, and through its Centres
of Excellence funding scheme Grant No. 262633 “QuSpin.” J. L. also acknowledges Support from
Sigma2 - the National Infrastructure for High Performance
Computing and Data Storage in Norway, project NN9577K. J.W.A.R. acknowledge funding from the EPSRC Programme Grant “Superspin” (no. EP/N017242/1) and EPSRC International Network Grant “Oxide Superspin” (no. EP/P026311/1). Over the years I.\v{Z}. was supported by U.S. DOE, Office of Science BES, Award No. DE-SC0004890, U.S. ONR through Grants No. N000141712793 and MURI No. N000142212764 and U.S. NSF ECCS Grants No. 2130845,  No. 1810266, and U.S. AFOSR Grant No. FA9550-22-1-0349. N.B. was supported by supported by EPSRC through the New Investigator Grant EP/S016430/1. Nordita is funded in part by NordForsk.

\clearpage
\newpage
\maketitle{List of abbreviations}

\begin{itemize}
\item 1D = one-dimensional
\item 2D = two-dimensional
\item 2DEG = two-dimensional electron gas
\item AC = alternating current
\item BCS = Bardeen-Cooper-Schrieffer 
\item BTK = Blonder-Tinkham-Klapwijk
\item BIA = bulk inversion asymmetry
\item DOS = density of states
\item F = ferromagnet
\item FMR = ferromagnetic resonance
\item FFLO = Fulde-Ferrell-Larkin-Ovchinnikov
\item HM = heavy metal
\item ISHE = inverse spin Hall effect
\item IP = in-plane
\item IV = current-voltage
\item JJ = Josephson junction
\item LE-$\mu$SR = low-energy muon spin rotation 
\item MAAR = magnetoanisotropic Andreev reflection
\item MR = magnetoresistance
\item MZM = Majorana zero mode
\item NiFe, permalloy = Py
\item N = normal metal
\item N/F = normal metal/ferromagnet
\item OOP = out-of plane
\item QD=quantum dots
\item S = superconductor
\item SEM = scanning electron microscope
\item S/F = superconductor/ferromagnet
\item S/F/S = superconductor/ferromagnet/superconductor
\item S/HM = superconductor/heavy-metal
\item S/HM/F = superconductor/heavy-metal/ferromagnet
\item SIA = structure inversion asymmetry
\item SFQ = single flux quantum
\item SPOT = symmetries in the indices of spin, parity, orbit and time
\item SQUID = superconducting quantum interference device
\item SOC = spin-orbit coupling
\item SOC/S = spin-orbit coupled/superconductor
\item TAMR = tunneling anisotropic magnetoresistance
\item TMR = tunneling magnetoresistance
\item TRS = time reversal symmetry
\item vdW = van der Waals
\end{itemize}

\clearpage
\newpage
\bibliography{References}
\end{document}